\documentclass[twocolumn]{aastex631}

\newcommand{\overbar}[1]{\mkern 1.5mu\overline{\mkern-1.5mu#1\mkern-1.5mu}\mkern 1.5mu}

\usepackage{braket}
\usepackage{upgreek}
\usepackage{url}
\usepackage{comment}
\usepackage[]{listings}
\usepackage{graphicx}
\usepackage{subcaption}
\usepackage[colorinlistoftodos]{todonotes}
\usepackage[section]{placeins}
\usepackage{amssymb}

\usepackage{natbib}
\setcitestyle{square, comma, numbers,sort&compress, super}

\usepackage{colortbl}
\usepackage{amsmath}
\usepackage{gensymb}

\usepackage{booktabs, multirow} 
\usepackage{soul}

\usepackage{bm}
\usepackage{enumitem}
\definecolor{color1}{rgb}{0.6,0.,0.8}

\hypersetup{linkcolor=color1,citecolor=color1,filecolor=color1,urlcolor=color1}

\begin{document}

\title{The Atacama Cosmology Telescope: Machine Learning Driven Tools for Detecting Millimeter Sources in Timestream Pre-processing}

\correspondingauthor{Simran K. Nerval}
\email{simran.nerval@mail.utoronto.ca}

\author[0009-0006-0076-2613]{Simran K. Nerval}
\affiliation{David A. Dunlap Department of Astronomy \& Astrophysics, University of Toronto, 50 St. George St., Toronto, ON M5S 3H4, Canada} 
\affiliation{Dunlap Institute for Astronomy and Astrophysics, University of Toronto, 50 St. George St., Toronto, ON M5S 3H4, Canada}

\author[0009-0004-8314-2043]{Erika Hornecker} 
\affiliation{David A. Dunlap Department of Astronomy \& Astrophysics, University of Toronto, 50 St. George St., Toronto, ON M5S 3H4, Canada}

\author[0000-0002-1697-3080]{Yilun Guan}
\affiliation{Dunlap Institute for Astronomy and Astrophysics, University of Toronto, 50 St. George St., Toronto, ON M5S 3H4, Canada}

\author{Zeling Zhang}
\affiliation{Department of Computer Science, University of Toronto, 50 St. George St., Toronto, ON M5S 3H4, Canada}

\author[0000-0003-1690-6678]{Adam Hincks}
\affiliation{David A. Dunlap Department of Astronomy \& Astrophysics, University of Toronto, 50 St. George St., Toronto, ON M5S 3H4, Canada}
\affiliation{Specola Vaticana (Vatican Observatory), V-00120, Vatican City State}

\author[0000-0002-2840-9794]{Emily Biermann}
\affiliation{Los Alamos National Laboratory, Los Alamos, NM 87545, USA}

\author[0000-0003-2358-9949]{J.~Richard~Bond}
\affiliation{Canadian Institute for Theoretical Astrophysics, University of Toronto, 60 St. George St., Toronto, ON M5S 3H8, Canada}

\author[0000-0002-9711-9969]{Justin Clancy}
\affiliation{School of Physics, The University of Melbourne, Parkville, VIC 3010, Australia}

\author{Rolando D\"{u}nner}
\affiliation{Instituto de Astrof\'isica and Centro de Astro-Ingenier\'ia, Facultad de F\'isica, Pontificia Universidad Cat\'olica de Chile, Av. Vicu\~na Mackenna 4860, 7820436 Macul, Santiago, Chile}

\author[0000-0002-7145-1824]{Allen Foster}
\affiliation{Department of Physics, Jadwin Hall, Princeton University, Princeton, NJ 08544, USA}

\author[0000-0002-4765-3426]{Carlos Herv\'ias-Caimapo}
\affiliation{Instituto de Astrof\'isica and Centro de Astro-Ingenier\'ia, Facultad de F\'isica, Pontificia Universidad Cat\'olica de Chile, Av. Vicu\~na Mackenna 4860, 7820436 Macul, Santiago, Chile}

\author[0000-0002-0965-7864]{Ren\'{e}e Hlo\v{z}ek}
\affiliation{David A. Dunlap Department of Astronomy \& Astrophysics, University of Toronto, 50 St. George St., Toronto, ON M5S 3H4, Canada} 
\affiliation{Dunlap Institute for Astronomy and Astrophysics, University of Toronto, 50 St. George St., Toronto, ON M5S 3H4, Canada}

\author[0000-0002-5564-997X]{Thomas~W.~Morris}
\affiliation{Department of Physics, Yale University, New Haven, CT 06511, USA}
\affiliation{National Synchrotron Light Source II, Brookhaven National Laboratory, Upton, NY 11973, USA}

\author[0000-0002-4478-7111]{Sigurd N{\ae}ss}
\affiliation{Institute of Theoretical Astrophysics, University of Oslo, Sem Sælands vei 13, 0371 Oslo, Norway}

\author[0000-0003-1842-8104]{John Orlowski-Scherer}
\affiliation{Department of Physics and Astronomy, University of Pennsylvania, 209 South 33rd Street, Philadelphia, PA 19104, USA}

\author[0000-0002-8149-1352]{Crist\'obal Sif\'on}
\affiliation{Instituto de F\'isica, Pontificia Universidad Cat\'olica de Valpara\'iso, Casilla 4059, Valpara\'iso, Chile}

\author[0000-0001-5733-2717]{Jesse Treu}
\affiliation{Domain Associates, LLC, USA}

\begin{abstract}
We present a new pipeline utilizing machine learning for classifying short-duration features in raw time-ordered data (TOD) of cosmic microwave background survey observations. The pipeline, specifically designed for the Atacama Cosmology Telescope, works in conjunction with the previous TOD preprocessing techniques that employ statistical thresholding to indiscriminately remove all large spikes in the data, whether they are due to noise features, cosmic rays, or true astrophysical sources, in a process called ``data cuts". This has the undesirable effect of excising real astrophysical sources, including transients, from the data. The classification pipeline demonstrated in this work uses the output from these data cuts and is able to differentiate between electronic noise, cosmic rays, and point sources, enabling the removal of undesired signals while retaining true astrophysical signals during TOD preprocessing. We achieve an overall accuracy of 90\% in categorizing data spikes of different origin and, importantly, 94\% for identifying those caused by astrophysical sources. Our pipeline also measures the amplitude of any detected source seen more than once and produces a subminute-to-minute light curve, providing information on its short timescale variability. This automated pipeline for source detection and amplitude estimation will be particularly useful for upcoming surveys with large data volumes, such as the Simons Observatory.
\end{abstract}



\section{Introduction}  

Ground-based microwave astronomy has experienced a revolution in the past two decades, thanks to the deployment of large aperture telescopes that have thousands of detectors, increased their resolution, and advances in detector and read-out technology that have drastically improved their sensitivity while also allowing for simultaneous observation of multiple wavelength bands. For example, the Atacama Cosmology Telescope (ACT; \citealt{swetz/etal:2011,thornton/etal:2016,henderson/etal:2016}) observed the cosmic microwave background (CMB) with $\sim$\,3000 detectors across up to three frequency bands, covering six bands from ${\sim}$30 to 270\,GHz during its operation from 2007 to 2022.\footnote{Jupyter notebooks outlining how to use the latest ACT data products (ACT DR6) can be seen at \url{https://github.com/ACTCollaboration/DR6_Notebooks}, including a glitch classification notebook outlining the methods discussed in this paper.} ACT will be succeeded by the Simons Observatory (SO), which is initially set to deploy over 60,000 detectors split between large- and small-aperture telescopes across six frequency bands from ${\sim}$30 to 280\,GHz \citep{ade/etal:2019,so_collaboration:2025}. During observations, detectors, such as transition edge sensors, record the incident power at a cadence of a few hundred hertz as the telescope scans in azimuth at a constant elevation. In the ACT survey, the data timestreams of all detectors are packaged into files known as ``time-ordered data" (TOD), each containing roughly 11 minutes worth of data. Using knowledge of the telescope pointing, which is recorded during data collection, the TODs are later processed into microwave sky maps using maximum likelihood map-making \citep{Dunner2013ApJ...762...10D, Choi2020JCAP...12..045C, ACT:Aiola:2020}.

In addition to using the CMB directly as a probe of cosmological physics, an emerging use of CMB experiments is the study of the time-varying microwave sky. Wide-area surveys contain thousands of radio sources; the majority of them are blazars that emit microwave light that changes in flux over days/weeks. The South Pole Telescope (SPT) collaboration \citep{2011Carlstrom} published a pilot study of one blazar (PKS 2326-502) observed in the microwave bands \citep{2023Hood}. Surveys like SO will provide much longer baselines of observations of such objects: the field of mm-transient studies is expected to grow significantly in the coming decade. Similarly in the realm of planetary science, ACT and SPT have also characterized the millimeter flux of asteroids \citep{chichura/etal:2022,orlowski-sherer/etal:2024}, and ACT constrained the existence of a ``Planet 9" in our Solar System \citep{2021planet9}. Finally, ACT and SPT have successfully measured dozens of transients in their data that are mainly flaring stars, with the majority consisting of short-duration (minutes- to hours-long) stellar flares \citep{whitehorn/etal:2016,naess/etal:2021,li/etal:2023,guns/etal:2021,tandoi/etal:2024,Biermann:2024emf}. Additionally, as part of a blind transient search, ACT has detected classical nova YZ Ret during its outburst, making it only the second millimeter observation of this kind \citep{Biermann:2024emf}. ACT also published a targeted search for extragalactic transients that included gamma ray bursts (GRBs), tidal disruption events (TDEs), and supernovae (SN), placing upper-limits on the microwave flux of the target objects \citep{hervias-caimapo/etal:2024}. More sensitive data from the upcoming SPT, SO, and CMB-S4 \citep{2016cmbs4} surveys will produce detections in the millimeter of extragalactic transients, specifically GRBs, which will provide valuable information on the emission processes of these powerful explosions \citep{eftekhari/etal:2022,so_collaboration:2025}. 

CMB surveys are designed to measure the power of faint microwave radiation imprinted on all angular scales across the sky. However, the data collected by ground-based CMB telescope like ACT are dominated by microwave emission from water vapor in the atmosphere and are additionally contaminated from, e.g., instrumental noise and cosmic rays colliding with the detectors. While instrumental and atmospheric noise can be modeled in the microwave map-making process to produce unbiased, maximum likelihood maps \citep{Dunner2013ApJ...762...10D}, the mapmaker is specifically vulnerable to biases from short, high-amplitude, non-Gaussian data bursts in the TOD, and therefore depends on these bursts being excised from the TOD before map-making. Such transient events can be caused by sporadic pathologies from the electronic readout system, cosmic rays, electromagnetic interference, radio frequency interference modulated by telescope motion, ionizing radiation, digitization artifacts, etc \citep{PlanckCR, digitization, RFI}. Cosmic rays, for example, sometimes produce rapid spikes in the TODs of single detectors and may sometimes also warm up a region of the detector polarization array (PA) near the impact, thereby inducing a signal spike in multiple detectors (see \cite{PlanckCR} for a study on cosmic rays for Planck High Frequency Instrument). In this paper, we refer to all of these transient phenomena as ``glitches''. Bright astrophysical point sources can also cause high-amplitude, non-Gaussian spikes in the TOD and may be mistaken for a glitch. This may be avoided if the coordinate of the source is known before pre-processing. However, this also means that any point source or astrophysical transient that is not already known in advance of map-making can be removed from the data set entirely. We note that while the specific phenomena we describe here are commonly encountered in ACT timestreams, other transient sources may manifest differently in different telescopes depending on their readout schemes and observing conditions. 

It is therefore imperative when preprocessing TODs from CMB telescopes to excise as many glitches as possible while still retaining as many true astrophysical transients as feasible. As the volume of data observed by current and future experiments increases, traditional techniques that rely on human inspection of a significant fraction of candidate glitches in order to tune statistically based thresholding algorithms will rapidly become unwieldy. Improvements in the data-excision algorithms are needed, both to better automate the process and to avoid cutting real signals. In this paper, we describe a new method utilizing machine learning that achieves these objectives utilizing the outputs of the current glitch finder.

After reviewing in Section~\ref{sec:state_of_act_cuts} the ``cuts process'' used by ACT to cut out potential glitches from the data, we describe our data preparation in Section~\ref{sec:data_preparation} and introduce the supervised machine learning algorithm in Section~\ref{sec:classification_algorithm}. We explain the summary statistics derived from the TODs that are fed into the classifier in Section~\ref{sec:classification_algorithm} and how we prepare our training and test sets in Section~\ref{sec:training_set}. We test the classification method on simulated stellar flares which are discussed in Section~\ref{sec:sims}, summarize and discuss our results in Section~\ref{sec:results}, and present our conclusions and suggestions for future work in Section~\ref{sec:conclusions}.

\section{Current State of ACT Data Cuts}\label{sec:state_of_act_cuts}

Before raw ACT TODs are processed into CMB maps by the map-making pipeline, portions of the TODs suspected to be glitches or contaminated with noise are excised from the data in a process known as ``data cuts''. The data cuts algorithm in ACT is described in \cite{ACT:Aiola:2020} and in detail in \cite{Dunner2013ApJ...762...10D}; however, we summarize some key steps and properties of the cuts algorithm below.

First, the data cuts algorithm can discard entire TODs based on the level of precipitable water vapor in the atmosphere during the TOD, or the number of faulty detectors operating over that time period. Additionally, we flag individual detectors within each TOD based on various statistical metrics computed on the TOD. A detector may become faulty for various reasons, such as saturation from optical and electric power, an instability caused by insufficient heat capacity, slow time constants (which describes how quickly the detector changes in response to a signal) from fabrication and wiring problems, or failures in the detector readout system. Current methods for producing individual detector cuts utilize a set of parameters to identify and exclude such detectors with anomalous behavior. These anomalies often manifest as abnormally high or low noise levels, poor optical coupling, or excessive high-frequency, non-Gaussian noise above around 10~Hz, where photon shot noise with a white spectrum is expected to dominate. Detectors that have outlier values of the  statistics for these parameters are flagged and rejected from map-making.

Moreover, short, high signal-to-noise (SNR) spikes on millisecond (ms) time scales are also prevalent in the detector timestreams, caused by glitches in the readout system and cosmic rays, among other factors. ACT employs a glitch finder that combines high-pass and Gaussian filters to identify samples affected by glitches, defined as those with $\mathrm{SNR} \geq 10$ in the high-pass-filtered TOD. This method is effective because ACT TODs are predominantly noise dominated. When a glitch is flagged, a buffer of 200 time samples on either end of the glitch is identified, and all samples from the detector within this buffer interval are rejected from the map-making process.

While this glitch finder is effective at removing glitches due to non-astrophysical signals, it also excludes bright astronomical sources that appear as spikes in the timestream as the telescope scans across them. Given the ACT scanning speed ($\sim1\degree\mathrm{s}^{-1}$) and beam size $\theta_\mathrm{FWHM} \simeq 1.4' $ at 150~GHz, astrophysical source spikes typically last around 20--30\,ms. In the standard map-making pipeline, a catalog of bright astrophysical sources is curated and passed to the glitch finder; samples in the TOD that lie within $3'$ of a known source from the catalog are masked before running the map-maker. This approach ensures that known bright astrophysical sources will not be cut, but it necessitates multi-pass TOD processing and, more importantly, may fail when a previously unknown source of transient nature appears in the observing field, such as those reported in \citet{naess/etal:2021} and \citet{li/etal:2023}. If these signals are bright enough, they might be erroneously removed by the glitch finder, resulting in missed opportunities to discover new sources of astrophysical interest.

Different types of glitches typically manifest differently in the timestream and across the focal plane. For instance, cosmic rays often simultaneously affect detectors in close proximity to each other. In contrast, a signal from a bright astrophysical point source is extended on the focal plane given the finite beam-shape. Point source signals are thus extended in the timestream as well and appear at different times for different detectors depending on their locations on the focal plane. As the telescope scans the sky at a constant elevation, detectors that see the same source fall along a constant-elevation line on the focal plane during each scan. The distinct spatial and temporal characteristics of different glitches can in principle be used to differentiate them. The spatio-temporal information of the TOD data has also been successfully used to model the bulk motion of the atmosphere \citep{timelag}.

In this work, we demonstrate that by combining both spatial and temporal information for different classes of glitches, it is possible to effectively distinguish between them with the assistance of machine learning. This approach provides insights into the physical origins of each glitch, and offers a promising method for preserving and identifying astrophysically interesting transient signals in the TODs in a fully automated method without the need for visual inspection by human experts. This property of the glitch classification will have growing importance as the data volume scales up in the coming decade with surveys like SO.

\section{Data Preparation}\label{sec:data_preparation}

Within the methodology we present here, the first step is to calibrate and perform data cuts on each TOD, following the same steps to identify and excise glitches as outlined in Section~\ref{sec:state_of_act_cuts}. Since each physical event that triggers glitches in a TOD may affect multiple detectors at slightly different times, we group together glitches that occur in different detectors within 200 time samples of each other  (ACT has a sampling rate ranging from 300 - 400 Hz depending on the detector array). For example, if glitch 1 and glitch 2 occur within 200 samples of each other, and glitch 3 and glitch 2 occur within 200 samples of each other but glitch 1 does not, all three glitches are grouped together. We extract the samples from the impacted detectors within this joint interval and call that subset of the timestream a \textit{snippet}. Snippets thus group together samples from multiple detectors that are potentially affected by the same physical event, and can be analyzed to reveal the spatial and temporal structure of that physical event.

\begin{figure}
    \centering
    \begin{subfigure}[b]{0.4\textwidth}
   \includegraphics[width=\textwidth]{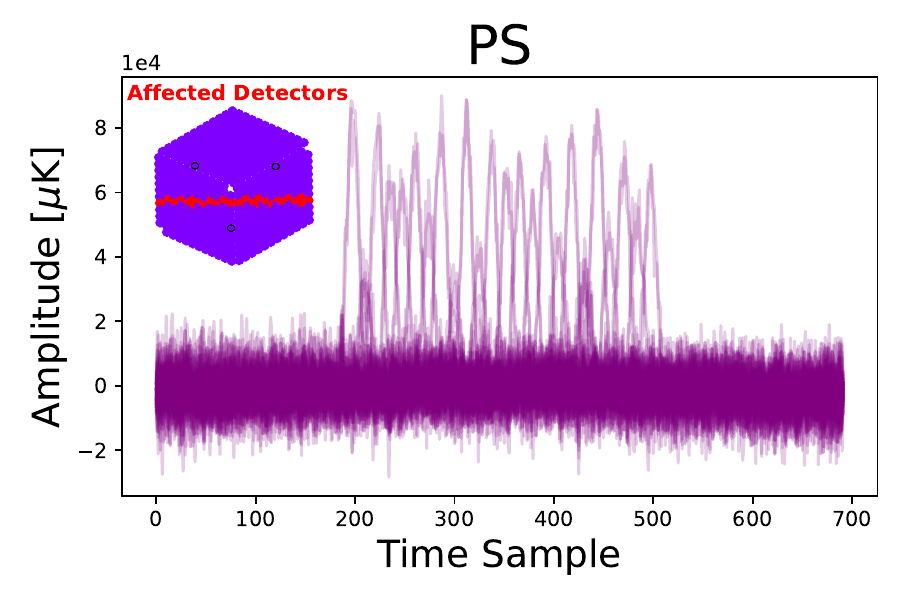}
	\end{subfigure}
 \begin{subfigure}[b]{0.4\textwidth}
   \includegraphics[width=\textwidth]{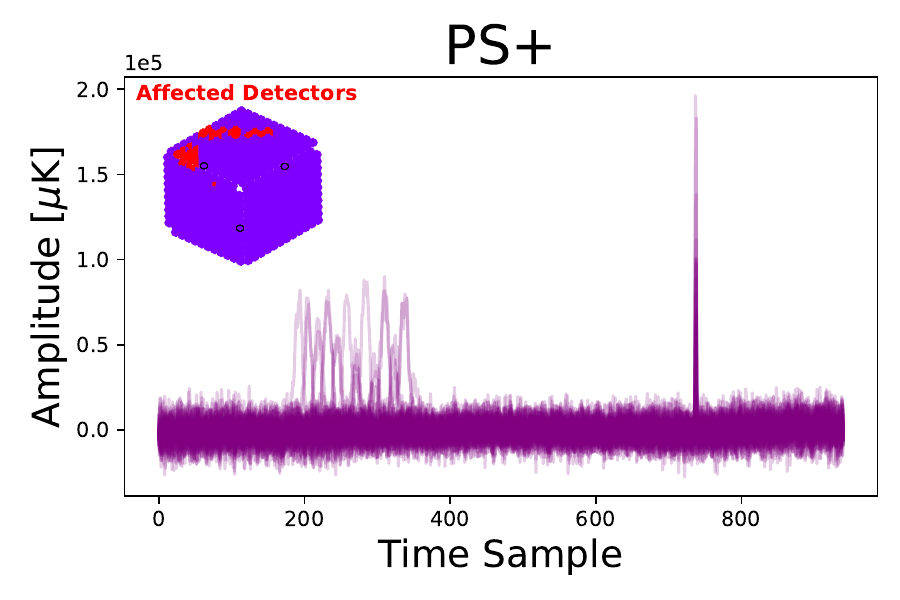}
	\end{subfigure}
 \begin{subfigure}[b]{0.4\textwidth}
   \includegraphics[width=\textwidth]{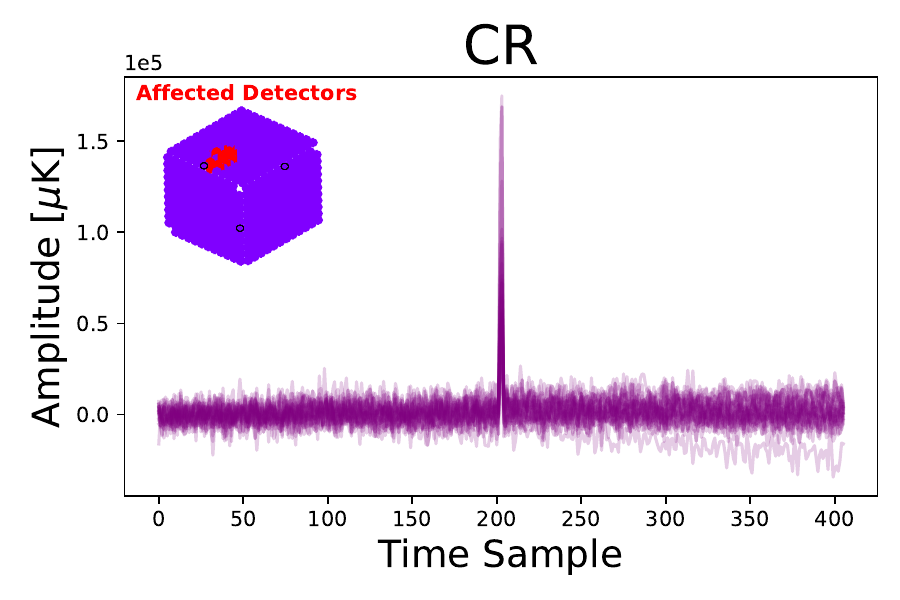}
	\end{subfigure}
 \begin{subfigure}[b]{0.4\textwidth}
\includegraphics[width=\textwidth]{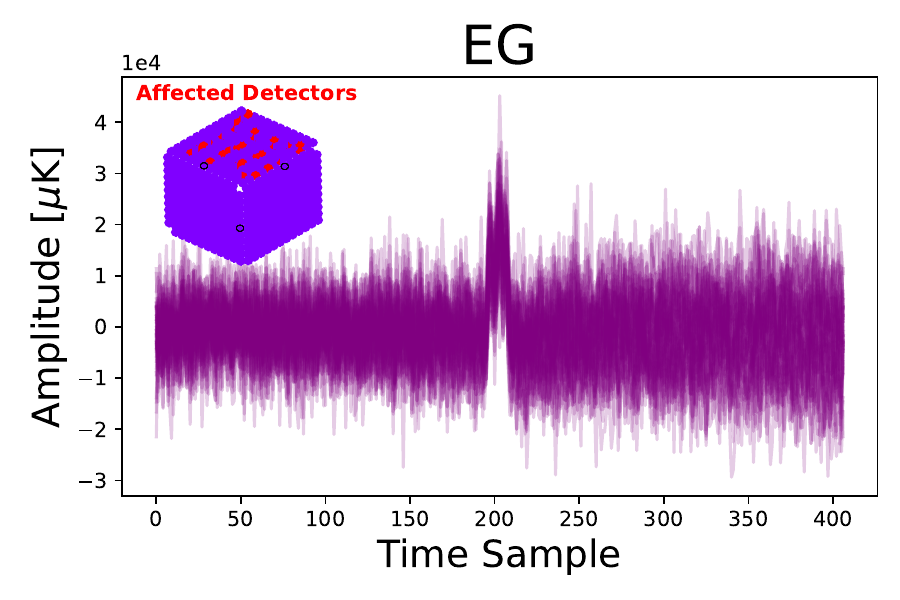}
	\end{subfigure}
\caption{The four categories we use in our classifications. From top to bottom we show example detrended TODs for a point source (PS), a point source with another coincident glitch (which is a cosmic ray in this case, labeled as PS+), a cosmic ray (CR), and an electronic glitch (EG). We show the timestreams for all affected detectors, along with the focal plane diagrams in the upper left of each panel, which show the affected detectors in red.}
 \label{fig:categories}
\end{figure}

Figure \ref{fig:categories} shows snippets for the four categories we define in this work to generally summarize the phenomena we see in the glitches for classification: point sources (PS), point sources with another coincident glitch (abbreviated as PS+), cosmic rays (CR), and electronic glitches (EG). In the figure, the example snippet for the PS+ contains a point source together with a cosmic ray. There is a large variety of electronic glitches that are observed in our data (all labeled as EG): for instance, in the bottom panel of Figure~\ref{fig:categories}, a diagonal ``striping'' is seen on the focal plane. These detectors share a common read-out line in the time-domain multiplexed system of our multi-channel electronics read-out system \citep{battistelli/etal:2008}. We characterize both the TODs and layout of the affected detectors across the focal plane of the snippets using the summary statistics described in the following section and use them as the features for classification.

\section{Description of Classification Features}\label{sec:classification_algorithm}
 
\begin{figure*}
\centering
\includegraphics[scale=.19]{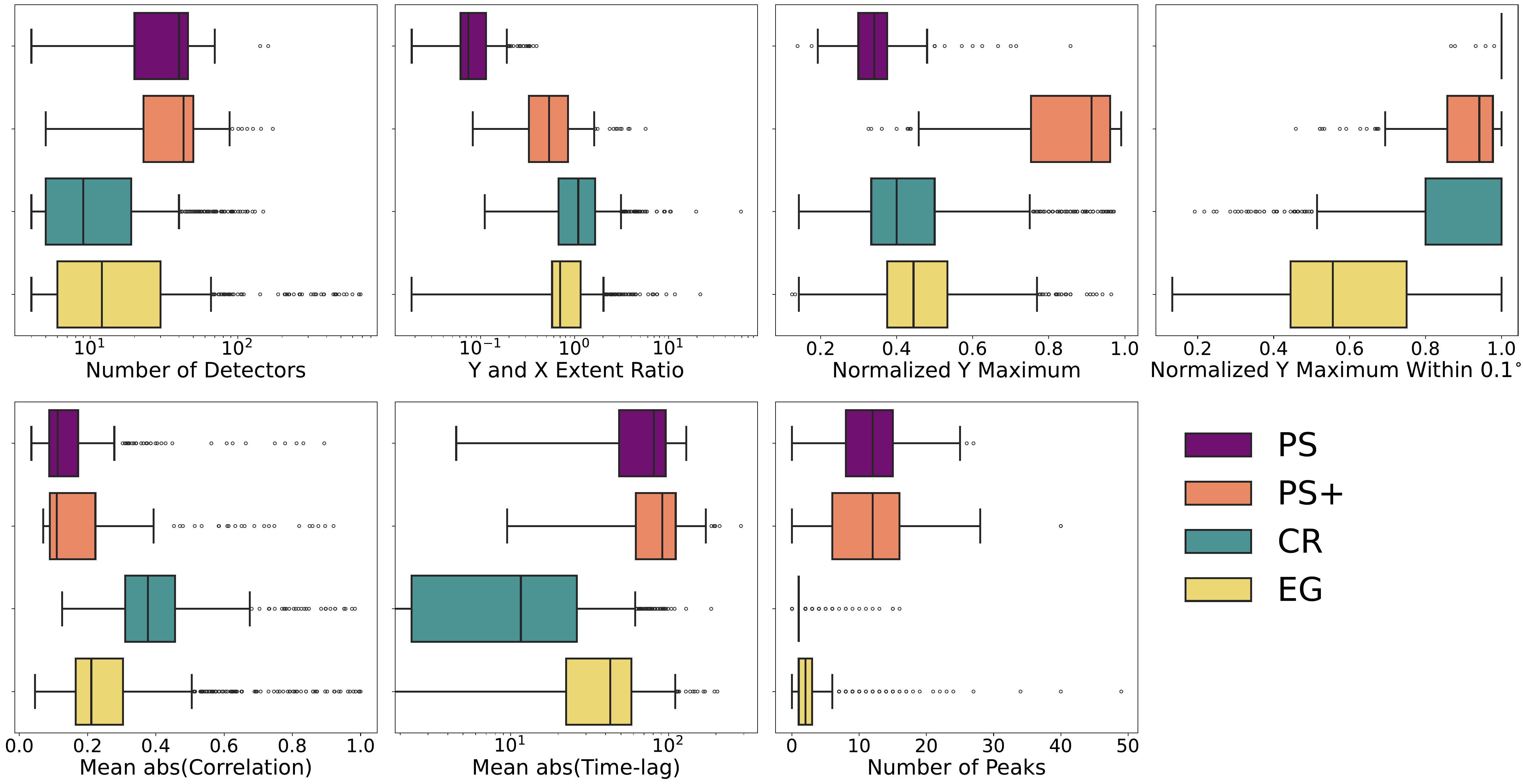}
\caption{The distributions of the summary statistics of the manually labeled training set used as the features in the random forest classifier for the categories we consider here: point sources (PS), point sources with another coincident glitch (PS+), cosmic rays (CR), and electronic glitches (EG). The boxes represent the first to third quartile with the median represented by the line in the box, the whiskers extend to 1.5 times the interquartile range, and outliers are represented with dots. The ``$Y$ and $X$ Extent Ratio" and ``Normalized $Y$ Maximum Within 0.1$\degree$" are the most useful to identify a PS, and the ``Normalized $Y$ Maximum" aids in separating a PS+ objects from a point source. Objects labeled as CR are best differentiated by the correlation, time-lag, and number of peaks. Electronic glitches have a variety of causes which results in different statistics, but in general, the normalized $Y$ maximum within 0.1$\degree$ statistic is lower than the other categories due to the affected detectors either generally being randomly distributed or following read-out lines. See Sections \ref{ssec: num dets}--\ref{ssec: num peaks} for definitions of each summary statistic.}\label{fig: dists}
\end{figure*}

We use a random forest machine learning algorithm, which can be thought of as a collection of ``decision trees'' that classify data based on the input features \citep{randomforest}. In order to differentiate the glitches produced by the current ACT glitch finder, we extract seven summary statistics per snippet from the timeseries and detector positions as features for the random forest, which can be further grouped into five distinct subtypes:
\begin{enumerate}
    \item The number of detectors affected by the glitch (Section~\ref{ssec: num dets}).
    \item Summaries of the distribution of these detectors across the focal plane ($Y$ and $X$ extent ratio, normalized $Y$ maximum, and normalized $Y$ maximum within 0.1$\degree$; Section~\ref{subsec:distributions}).
    \item The correlation of the signal (Section~\ref{ssec:correlation}).
    \item The time lag between detectors (Section~\ref{ssec:time_lag}).
    \item The number of peaks in the TODs (Section~\ref{ssec: num peaks}).
\end{enumerate}
Figure~\ref{fig: dists} shows the distributions of the seven statistics, described in the subsections that follow, falling in each of the five types enumerated above. Alternative unsupervised methods, that utilize focal plane images or the entirety of the temporal data for example, could be explored in future work. However, the goal here is to develop a computationally light method that could run on-site for future telescopes for real-time transient detection.

\subsection{Number of Detectors}\label{ssec: num dets}

\subsection{Detector Distributions Across the Focal Plane}
\label{subsec:distributions}

Multiple summary statistics are used to help characterize the affected detector distributions across the focal plane. Each detector has a unique $X$- and $Y$-coordinate in degrees on the focal plane. In order to compute useful statistics, the focal plane distributions are collapsed along the $X$- and $Y$-axes to create histograms summarizing these distributions. We would expect that stationary astrophysical sources to affect detectors along the scan direction which corresponds to the $X$-axis and the $Y$-coordinate distribution to fall within a few beams. Example focal planes for the glitch categories can be seen in Figure~\ref{fig:categories} and example $Y$ histograms for a PS and a PS+ are shown in Figure~\ref{fig:y dists}. 

\subsubsection{The $Y$ and $X$ Extent Ratio}
One of the statistics that can be derived from the focal plane is the ratio of the ranges in the $X$ and $Y$ directions, which is computed as:
\begin{equation}
    \text{ratio of ranges} = \frac{Y_{\text{max}} - Y_{\text{min}}}{X_{\text{max}} - X_{\text{min}}},
\end{equation} where $Y/X_{\text{max} (\text{min})}$ are the maximum and minimum values of the $Y/X$ distributions of the affected detectors respectively.

Since many detectors will observe a PS across the focal plane in a horizontal line as the telescope scans in azimuth, we expect a large spread in the $X$ coordinate but only a small range of affected detectors in the $Y$ direction, since the celestial sphere barely rotates during a single scan. Therefore, there should generally be a small ratio of $Y$ extent compared to the $X$ extent/range for a point source. A cosmic ray will hit a cluster of detectors, which results in a similar spread in the $Y$ and $X$ directions, and we therefore generally expect the ratio to have a value close to unity for CR objects. However, this is not always the case. For example, if multiple cosmic rays hit the focal plane at once, the ratio will deviate from unity, but the $Y$ and $X$ extent ratio will typically remain larger than that of a point source. The expected extent ratio also varies quite significantly for the EG category depending on the type of electronic glitch, however the ratio is generally larger than that for a point source. For a PS+, we expect a variety of deviations that differ from the extent ratio expected for a point source alone, depending on the location of the other glitch with respect to the location of the source on the focal plane. If the glitch occurs near the source, we similarly expect a small ratio. Conversely, if the glitch occurs far away on the focal plane, the ratio will be closer to one as it will have a large $Y$ range and the $X$ range is generally close to the full $X$ extent of the focal plane for a point source.

\subsubsection{Normalized $Y$ Maximum}

To derive this statistic, we compute the histogram of the distribution of the detectors in the $Y$ direction of the focal plane for a given glitch. The histogram bin with the largest number of detectors and the adjacent bins are added together and then normalized by the total number of detectors present in the glitch. We compute the histogram of detectors with 10 bins, resulting in the width of an individual bin in the histogram being small when the detectors are clustered and larger if there is spread in $Y$ positions. Figure \ref{fig:y dists} shows examples for both the PS and PS+ categories, where the detectors being considered are shown as coral compared to all the detectors in the snippet shown in purple. Note that in both examples, there is only one bin beside the maximum $Y$ bin, thus only two bins are used for the normalized $Y$ maximum statistic. For the PS+, this is because the maximum $Y$ bin is the first bin, whereas for the PS it is due to the detectors that would have been in the bin to the right of the maximum $Y$ being cut during the detectors cuts which happen before the glitch cuts. The PS object, has only 27\% of the detectors within the maximum and adjacent $Y$ bins, compared to 59\% of detectors within the maximum $Y$ and adjacent bins for the PS+ object.

This statistic is most useful for recognizing a PS+ object, as there tend to be numerous detectors that observe the point source itself compared to those that see only the coincident glitch. Conversely, because point sources and cosmic rays are typically tightly clustered and have less scatter in the $Y$ direction, a 10-bin histogram of the detectors has a smaller individual bin width and generally does not contain a large fraction of the affected detectors. Objects classified as EG have a variety of distributions across the focal plane and are thus not well characterized by this statistic.
\vspace{1em}
\subsubsection{Normalized $Y$ Maximum Within 0.1$\degree$}

\begin{figure}
\centering
\includegraphics[scale=.4]{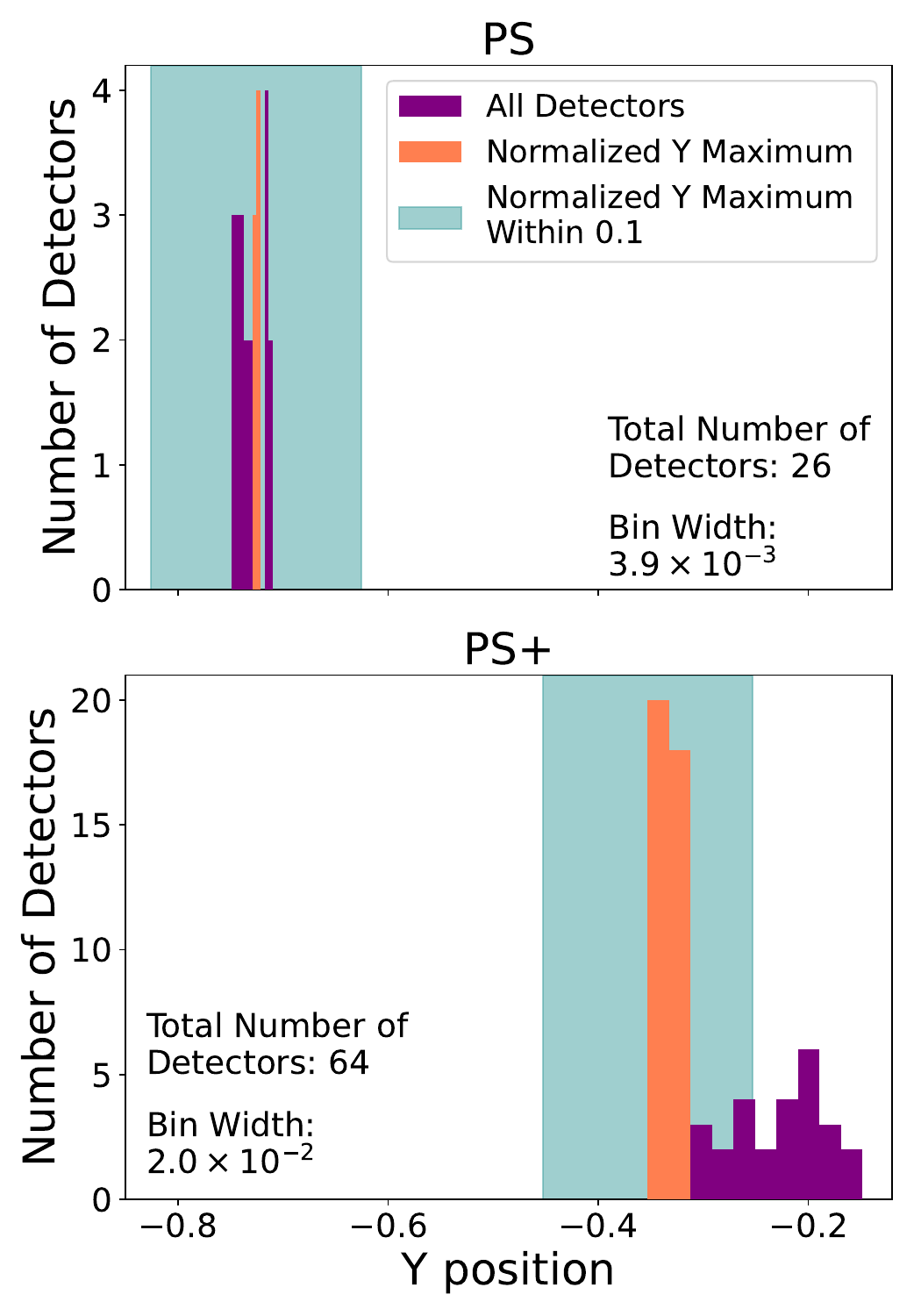}
\caption{Distribution of the detectors in the $Y$ direction of the focal plane for a PS (top) and PS+ (bottom). The purple histograms represent all the detectors that observe the glitch, while the coral histogram bins indicate the detectors being used to compute the normalized $Y$ maximum statistic, and the teal-shaded region includes all detectors that lie within 0.1$\degree$ of the maximum $Y$ bin, that define the normalized $Y$ maximum within 0.1$\degree$ statistic. Note that in both examples, there is only one bin beside the maximum $Y$ bin. The coral regions contain 27\% (two bins with a fraction of detectors given as (3+4)/26) of the total detectors that observe the PS (top row) and 59\% ((20 + 18)/64) of the total detectors that see the PS+ (bottom row). The teal regions contain 100\% of the detectors for the PS and only 67\% for the PS+. The distributions of this statistic for all glitch categories can be seen in Figure~\ref{fig: dists}.}\label{fig:y dists}
\end{figure}

In general, we expect the spread of affected detectors in the $Y$ direction to fall within 0.1$\degree$, or a few beams, for an astrophysical source. This is not true for a general electronic glitch or cosmic ray, which makes this metric useful for identifying astrophysical sources. We compute the fraction of detectors in the $Y$ direction of the focal plane that are within 0.1$\degree$ of the histogram bin that contains the maximum number of detectors. Figure~\ref{fig:y dists} shows examples for both the PS and PS+ categories for both the normalized $Y$ maximum and normalized $Y$ maximum within 0.1$\degree$ to compare these statistics.

Similar to the previously defined statistic, we expect a small $Y$ range of detectors to observe a point source as the telescope scans over the source. Therefore, the number of detectors within 0.1$\degree$ of the $Y$ position of the maximum histogram bin should be close to the total number of detectors in the snippet (and therefore our statistic normalized over the total number of detectors should be very close to unity, as shown in Figures~\ref{fig: dists} and \ref{fig:y dists}). This makes the statistic very useful for determining if a glitch is a point source. A cosmic ray will hit a small cluster of detectors, resulting in a small spread in the $Y$ direction. We still expect the fraction of the total number of detectors within 0.1$\degree$ of the $Y$ position to be close to unity, but there can be deviations from this if multiple cosmic rays hit the focal plane at once. The expected distribution of detectors varies quite significantly depending on the type of electronic glitch is occurring, but we find that the statistic is generally smaller for electronic glitches than for the PS class. For a PS+ object, the value of this statistic will differ from what is seen for an isolated source depending once again on where the additional glitch occurs on the focal plane relative to the source location: if the glitch occurs near the source, we similarly expect many detectors to be included, leading to a large fraction of the detectors to be within 0.1$\degree$ of the $Y$ maximum. Conversely, if the glitch occurs far away on the focal plane, this will result in a smaller fraction of detectors within 0.1$\degree$ of the $Y$ maximum and a smaller value for our statistic. This statistical metric is therefore useful in identifying PS+ objects where the coincident glitches are located elsewhere in the detector array and thus would be able to be easily separated in the snippets from the point source and for further analysis of the source.

\subsection{Correlation} \label{ssec:correlation}

\begin{figure}
    \centering
  \begin{subfigure}[b]{0.35\textwidth}
   \includegraphics[width=\textwidth]{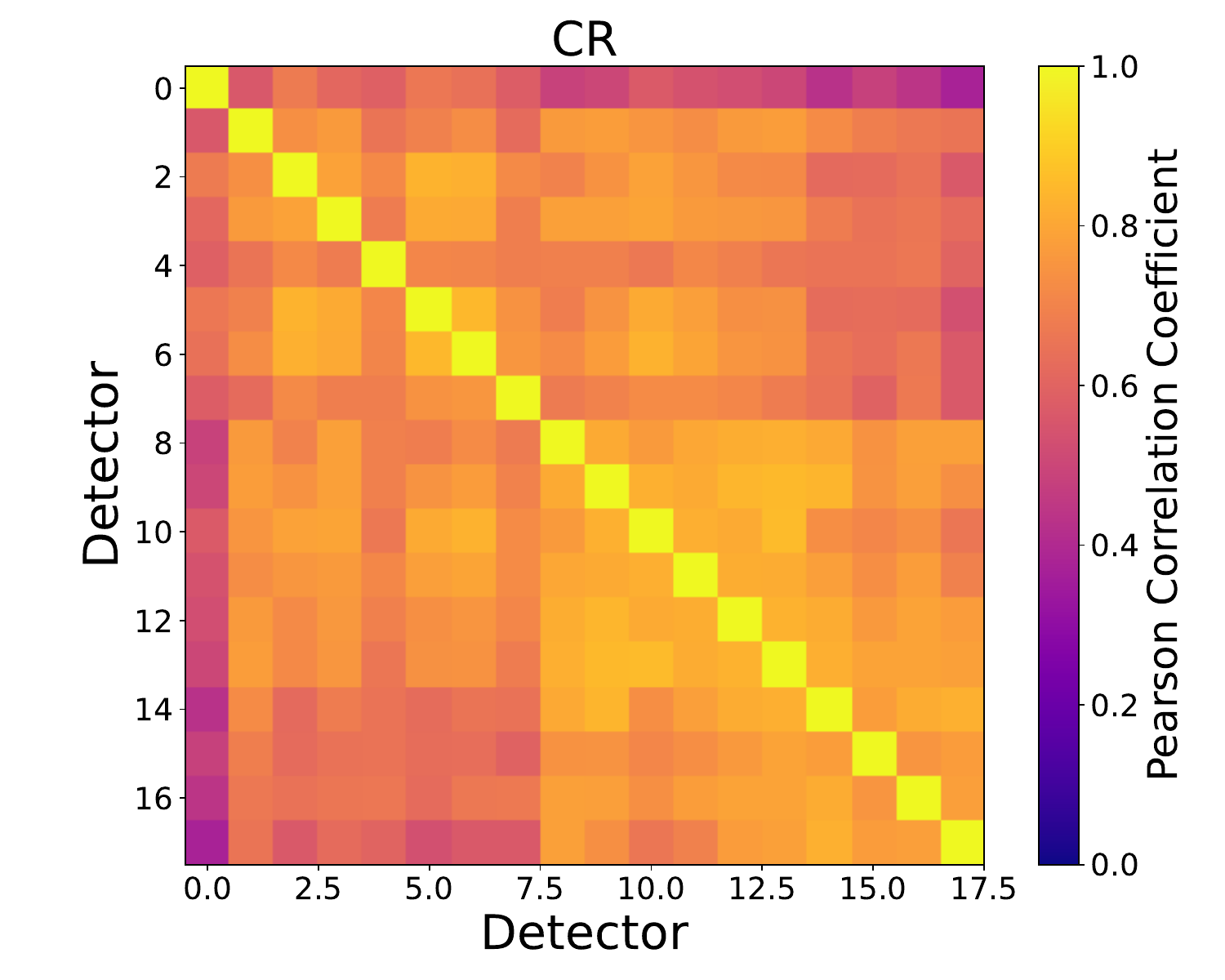}
	\end{subfigure}
  \begin{subfigure}[b]{0.35\textwidth}
   \includegraphics[width=\textwidth]{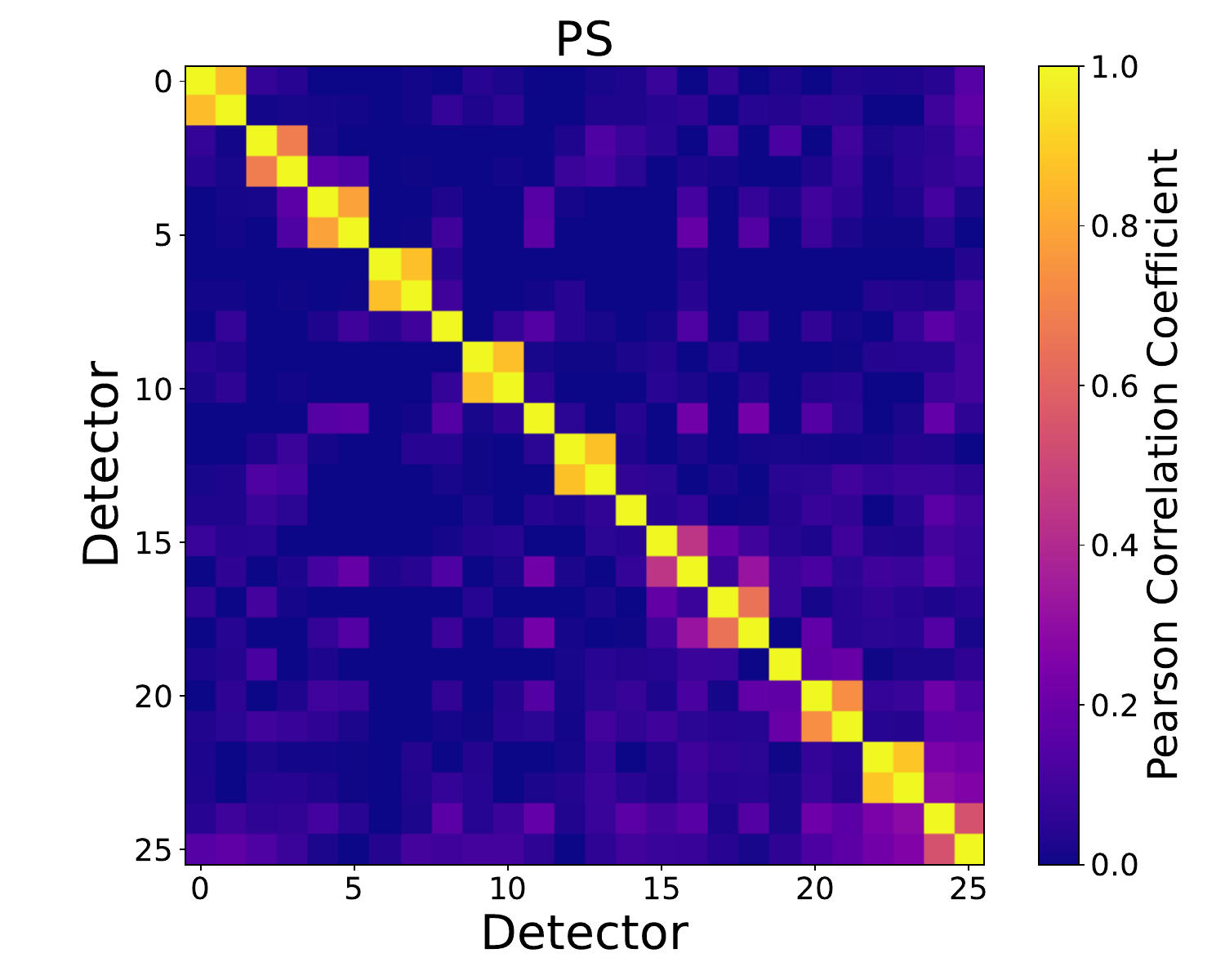}
   \end{subfigure}
    \caption{The correlation matrix between detectors, ordered by their $X$ position, affected by a cosmic ray (top) and a point source (bottom). Notice that for the point source, only detectors near each other have a high correlation, whereas the cosmic ray causes most detectors to be highly correlated. The distributions of this statistic for all glitch categories can be seen in Figure~\ref{fig: dists}. The mean of the absolute values of the correlation matrix values is used in our classification pipeline as a feature.}
    \label{fig:correlation}
\end{figure}

In order to determine how the incident flux is correlated across detectors (which can help distinguish between the PS and CR classes), we compute the Pearson correlation coefficient of each detector pair ordered by their $X$ position across the focal plane. We then use the mean of the absolute value of these correlation coefficients from all detector pairs in the snippet as a feature for our random forest. Given that detectors observe the point source sequentially as the telescope scans across the sky, we only expect a large value of this correlation coefficient between detectors that are either close to or beside the main detectors on the focal plane that observe the source. This scenario leads to a block diagonal correlation matrix of high values across the array when the detectors are ordered by their $X$ position (see Figure~\ref{fig:correlation} for an illustration) but a low overall degree of correlation across all detectors. For cosmic rays, we expect a high level of correlation across the array, because the detectors are all affected by the cosmic ray at the same time, shown in Figure~\ref{fig:correlation}. For electronic glitches that cause a strong signal that affects multiple detectors simultaneously, we expect a large value of the correlation coefficient during, e.g., a large-area heat-up. Conversely, if the glitch is similar to a random noise fluctuation, we would expect a low value of the correlation coefficient. For PS+ objects, if the additional glitch is a cosmic ray, we expect a slightly higher correlation than would be observed for a point source, as the cosmic ray portion of the TODs for each detector will overlap while affected by the cosmic ray. If the additional glitch is of electronic or other origin, the effect varies.

\subsection{Time-lag}\label{ssec:time_lag}

While the absolute value of the correlation described above indicates the spatial correlation between detectors that see a given glitch, the temporal correlation (or time-lag) is also a key statistic that we use in classification. The time-lag between detectors is the shift in time that maximizes the cross-correlation between the two detector TODs over the snippet. For two detectors $i$ and $j$, it is defined following \cite{timelag}:
\begin{equation} \label{eq: time lag}
    \tau = \text{max}\bigg[\text{DFT}^{-1}\Big[\text{DFT}[s_t^i]_f\cdot\overbar{\text{DFT}[s^j_t]}_f\Big]_{\tau}\bigg] , 
\end{equation}
\noindent where ``DFT" is the discrete Fourier transform, the overbar denotes the complex conjugate, and $s_t^{i/j}$ is the timestream of the $i/j^{\text{th}}$ detector as a function of time. We take the absolute value, since the sign of the shift (backwards or forwards) does not contain information relevant to the type of the signal. We compute this time-lag for all detector pairs in the snippet, and then use the mean of the absolute values to summarize the glitch. 

\begin{figure}
    \centering
  \begin{subfigure}[b]{0.35\textwidth}
   \includegraphics[width=\textwidth]{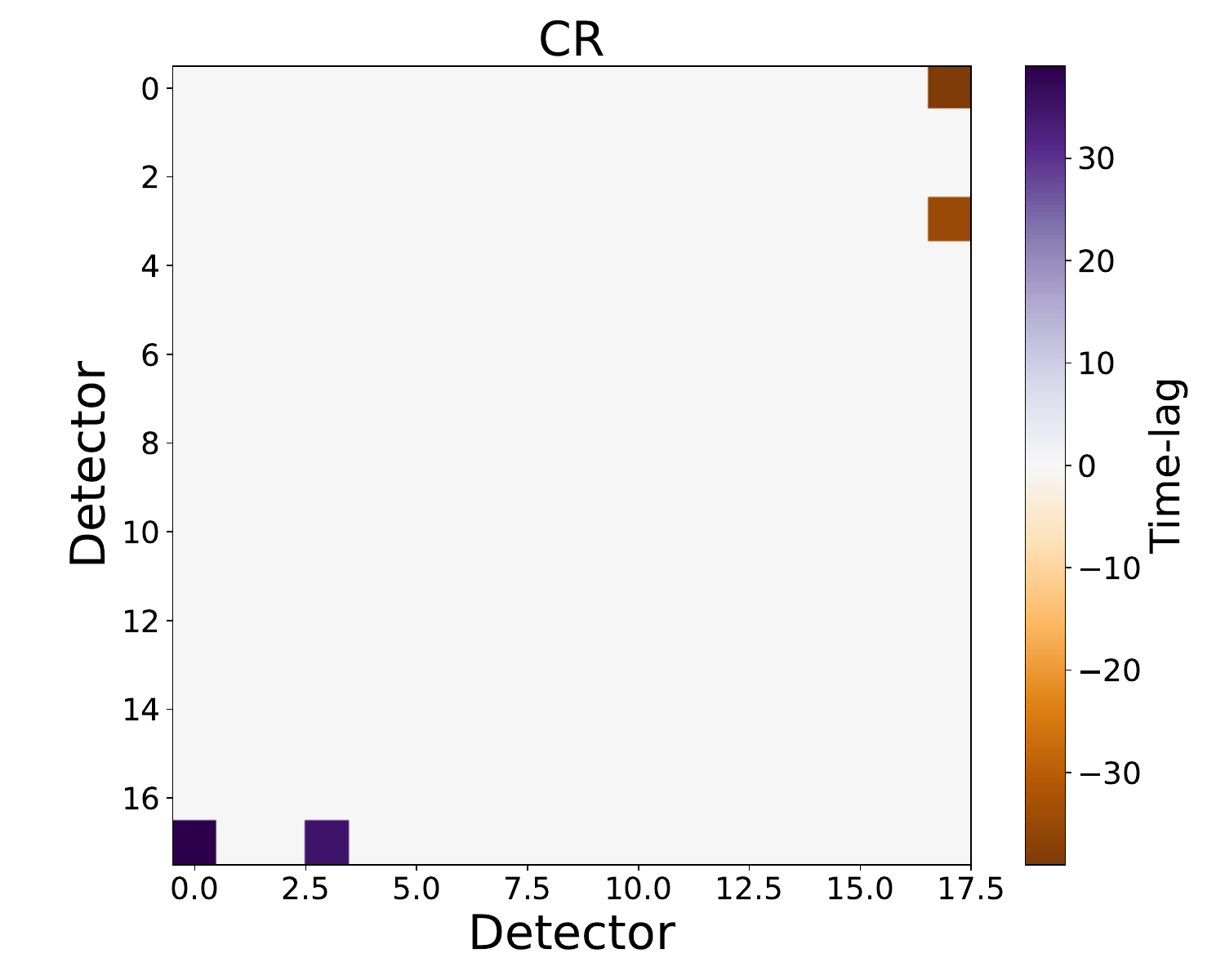}
	\end{subfigure}
  \begin{subfigure}[b]{0.35\textwidth}
   \includegraphics[width=\textwidth]{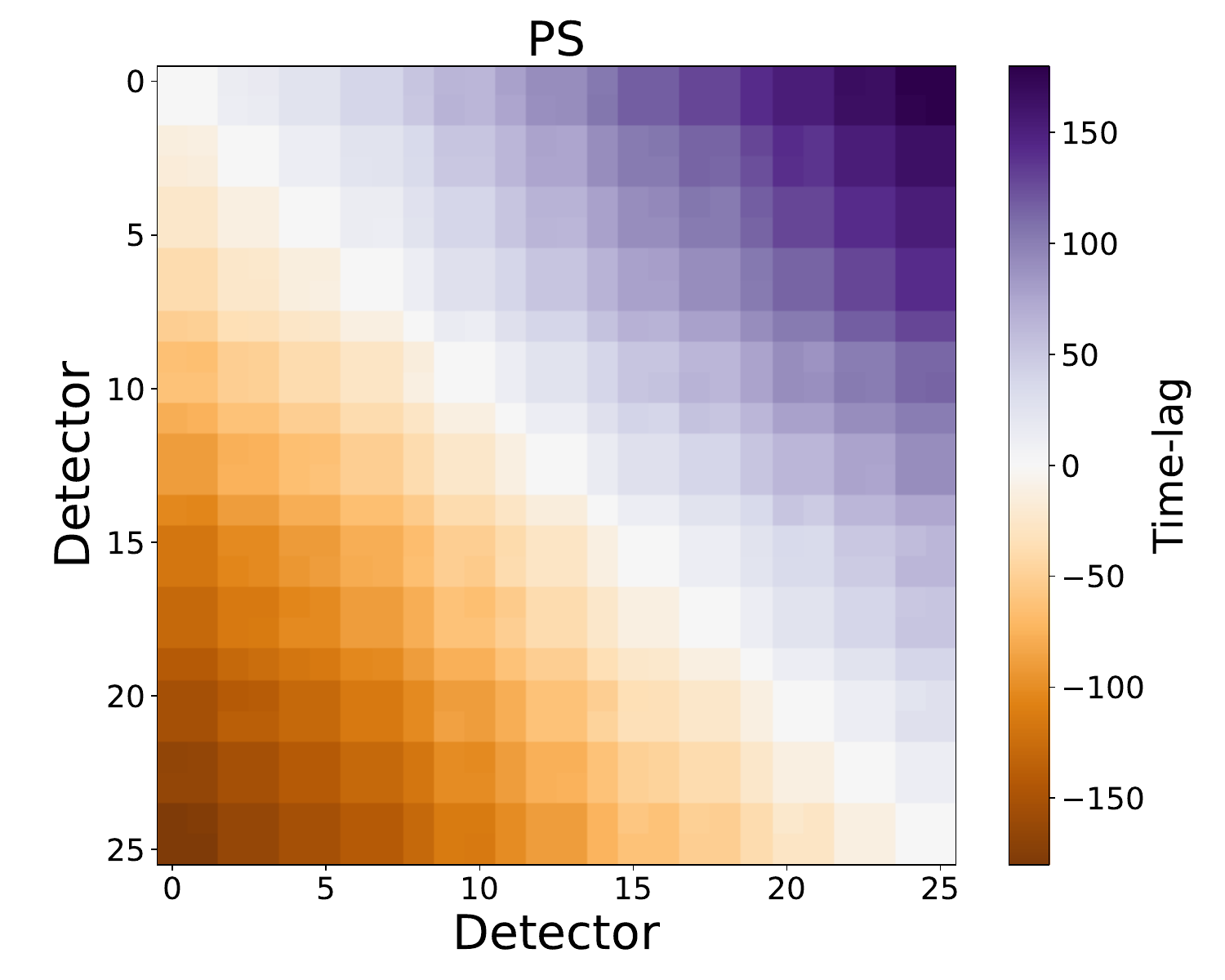}
   \end{subfigure}
    \caption{The time-lag matrix between detectors, ordered by their $X$ position, affected by a cosmic ray (top) and a point source (bottom). Notice that for the cosmic ray, most detectors have a time-lag of zero due to them being hit by the cosmic ray at the same time, whereas the point source has an increasing time-lag the further away the detectors pairs are from each other in the focal plane. The distributions of this statistic for all glitch categories can be seen in Figure \ref{fig: dists}. We use the mean of the absolute values of the time-lag values as our classification feature.}
    \label{fig:timelag}
\end{figure}

For a point source, each detector observes the source sequentially as the telescope scans across the sky. We therefore expect a time-lag that is equal to the separation in azimuth of the two detectors divided by the scanning speed. In practice, we can get different values of the time-lag based on how noisy the TOD is, especially for low amplitude sources. This can be used in future work to differentiate an astrophysical point source from a moving source, such as a satellite. Conversely, we expect a time-lag close to zero for cosmic rays, since the cosmic ray timescale is fast compared to the readout rate, which means that the digitized signatures reside essentially on top of each other in the detector timestream. This can be seen in Figure~\ref{fig:timelag}, where a time-lag matrix for detector pairs, ordered by their $X$ position, is shown for a cosmic ray on the top and a point source on the bottom. As expected, the time-lag is zero for almost every detector pair for the cosmic ray. For the EG category, it will cause varied time-lags depending on the nature of the glitch. Generally, something such as a focal plane heating event or another electronic glitch that affects all the detectors at the same time will result in a zero time-lag as the detectors experience the effect at the same time. Finally, for the PS+ class, if the other glitch is a cosmic ray, we expect a slightly smaller time-lag than just a point source as part of the TODs will be overlapping with a zero time-lag. If the other glitch is of an electronic or other origin the effect varies, but generally speaking it increases the time-lag, as we show in Figure~\ref{fig: dists}.
 
\subsection{Number of Peaks}\label{ssec: num peaks}

\begin{figure}
    \centering
    \begin{subfigure}[b]{0.4\textwidth}
   \includegraphics[width=\textwidth]{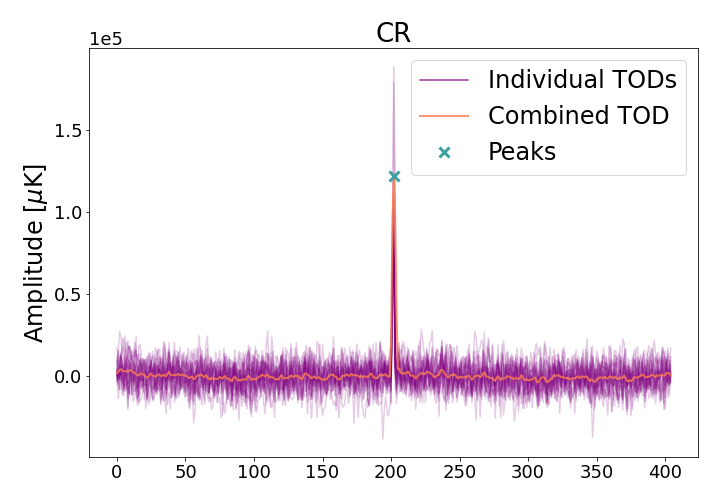}
	\end{subfigure}
 \begin{subfigure}[b]{0.4\textwidth}
   \includegraphics[width=\textwidth]{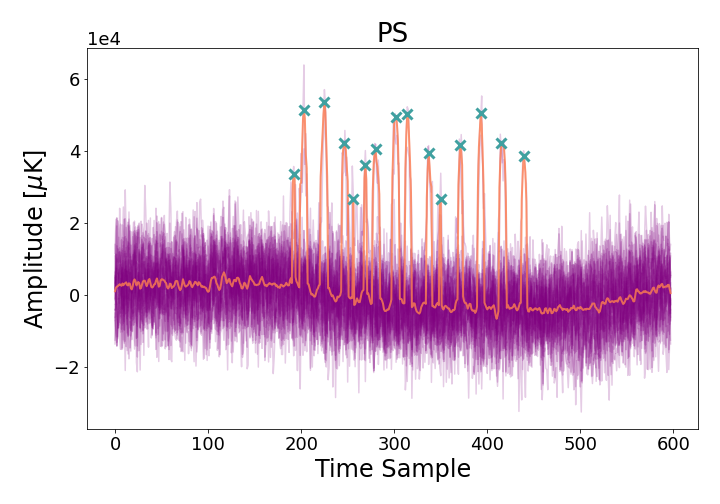}
	\end{subfigure}
    \caption{Example snippets for a cosmic ray (top) and a point source (bottom) showing the computation of the number of peaks in a TOD snippet combined over all detectors. The individual detector TODs are shown in purple, the combined TOD used to define the peaks is shown in coral, and the identified peaks are shown in teal. The distributions of this statistic for all glitch categories can be seen in Figure \ref{fig: dists}.}
 \label{fig:peaks}
\end{figure}

In a given snippet, some glitches, in particular glitches from PS objects and CR objects, will appear as ``peaks'' in the TOD. Thus, the number of signal peaks at unique times is used as one of our summary statistics, as it is particularly useful in discerning a cosmic ray from an electronic glitch. In order to compute the number of peaks in a snippet, we first smooth the TODs for the snippet using a boxcar smoothing kernel of width three time samples to remove small fluctuations due to random noise. We then combine the TODs from each detector $s^i_t$ into a single timestream $s_t$, as:

\begin{equation}
    s_t = \begin{cases}
        \max(s^i_t), & \text{if } \max(s^i_t) \geq 3\,\sigma_s \\
        \mu(s^i_t), & \text{otherwise},
    \end{cases}
\end{equation}

\noindent where $\max(s^i_t)$ and $\mu(s^i_t)$ indicate the maximum and mean of the all the samples at a given time $t$, respectively, and $\sigma_s$ is the standard deviation of all detectors and times in the snippets. Using this combined TOD, we use the Python \texttt{scipy.signal.find\_peaks} function to find the number and locations of the peaks for the TOD snippet. Examples for both a cosmic ray and a point source are shown in Figure~\ref{fig:peaks}. PS objects and combined PS+ objects generally have $\mathcal{O}(10)$ peaks, whereas most electronic glitches display fewer than five peaks, as can be seen in Figure~\ref{fig: dists}. Signals from cosmic rays generally only have one peak unless multiple cosmic rays hit simultaneously, which makes this statistic useful for breaking the degeneracy between them and electronic glitches.

\section{Training and Testing the Random Forest}\label{sec:training_set}

\subsection{Labeling of Objects in the Training and Test Sets} \label{ssec: labeling}

In order to create a large enough training set for the random forest classifier, we needed an efficient method to facilitate manual labeling of the snippets, via a graphical user interface (GUI).  

\begin{figure*}
\centering
\includegraphics[width=0.65\linewidth]{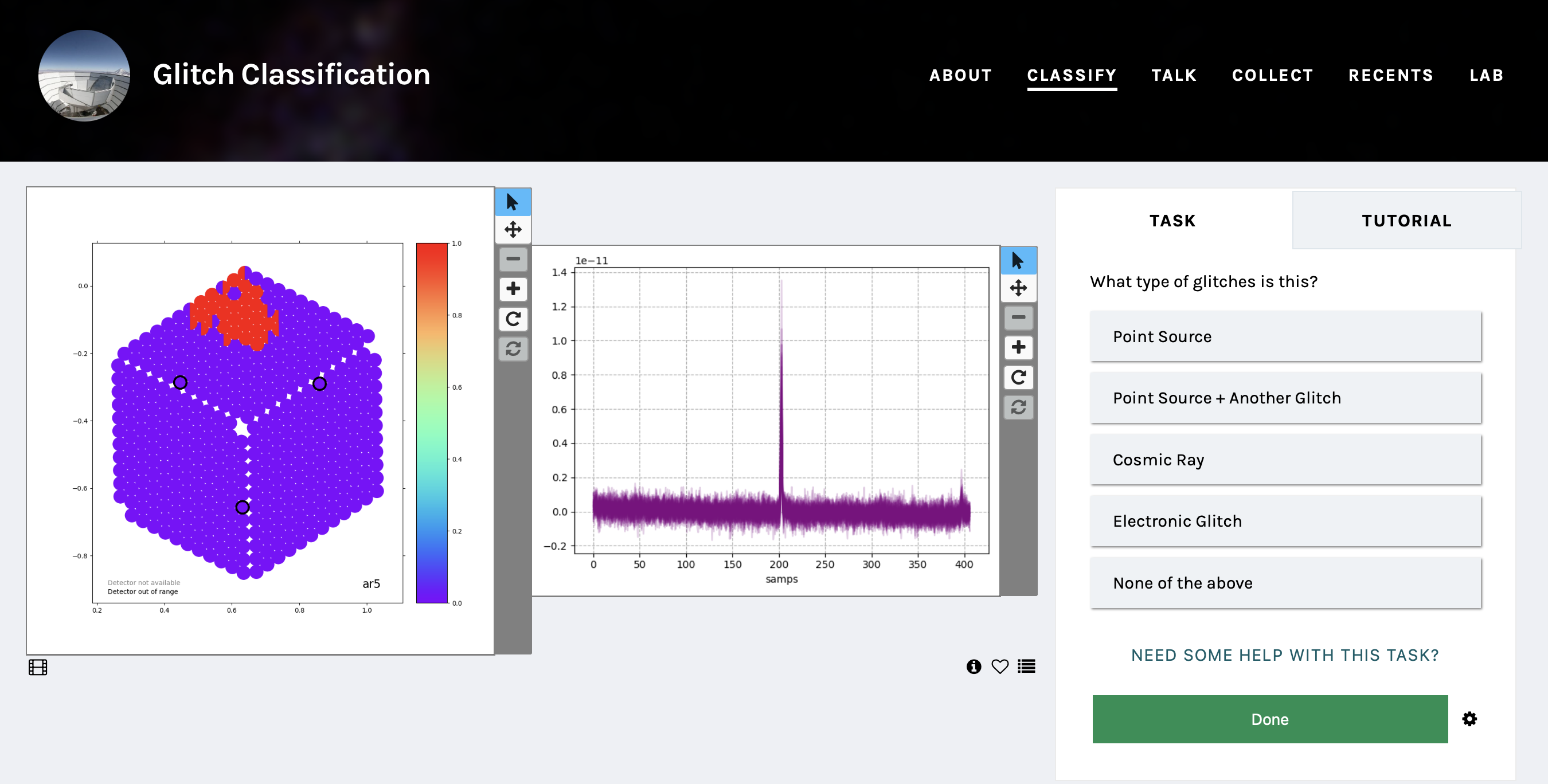}
\caption{Zooniverse interface for classifying the glitches. The displayed screen shows various elements of the tool, including a focal plane image on the left and a TOD on the right. Users can classify the nature of detected glitches by selecting options such as ``Point Sources," ``Cosmic Rays," or ``Electronic Glitch" from a drop-down menu. A tutorial section offers guidance on how to perform tasks. This tool assists users in analyzing and labeling anomalies in datasets, contributing to iterative learning and enhancement of the classification pipeline.}\label{fig:Zooniverse Interface}
\end{figure*}

We use the Python-driven platform developed by the Zooniverse collaboration\footnote{\href{www.zooniverse.org}{www.zooniverse.org}} to create the labeling GUI. This Zooniverse platform includes a well-established data pipeline and a user-friendly interface, as can be seen in Figure~\ref{fig:Zooniverse Interface}. 
The platform we employ was designed specifically for use by members of the ACT collaboration, to assist in both training and verification of the random forest.

To facilitate classifications by non-expert users, we created a user tutorial specific to ACT glitch data, with examples related to the key features of each type of glitch and notes on pitfalls in classifying glitches.\footnote{We separately included instructions on setting up a Zooniverse project from scratch, see \href{https://github.com/ZZBP/Zooniverse_Glitch_Classification_Code_Release}{online documentation here.}}

As shown in Figure~\ref{fig:Zooniverse Interface}, the user is shown a figure of the detector array, where the red detectors are those affected by the glitch. The user is also shown a figure of the TODs from all affected detectors superposed as a snippet, and is provided with five options for possible glitch labels (the four labels described in Section~\ref{sec:data_preparation} and an additional ``None of the above'' label).

The test and training data are then drawn from one larger sample of objects labeled in the manner described above. We describe the procedure for the separating this data set into a training set and a test set below.

\subsection{Forming the Training and Test Sets}

\begin{figure*}
\centering
\includegraphics[scale=.23]{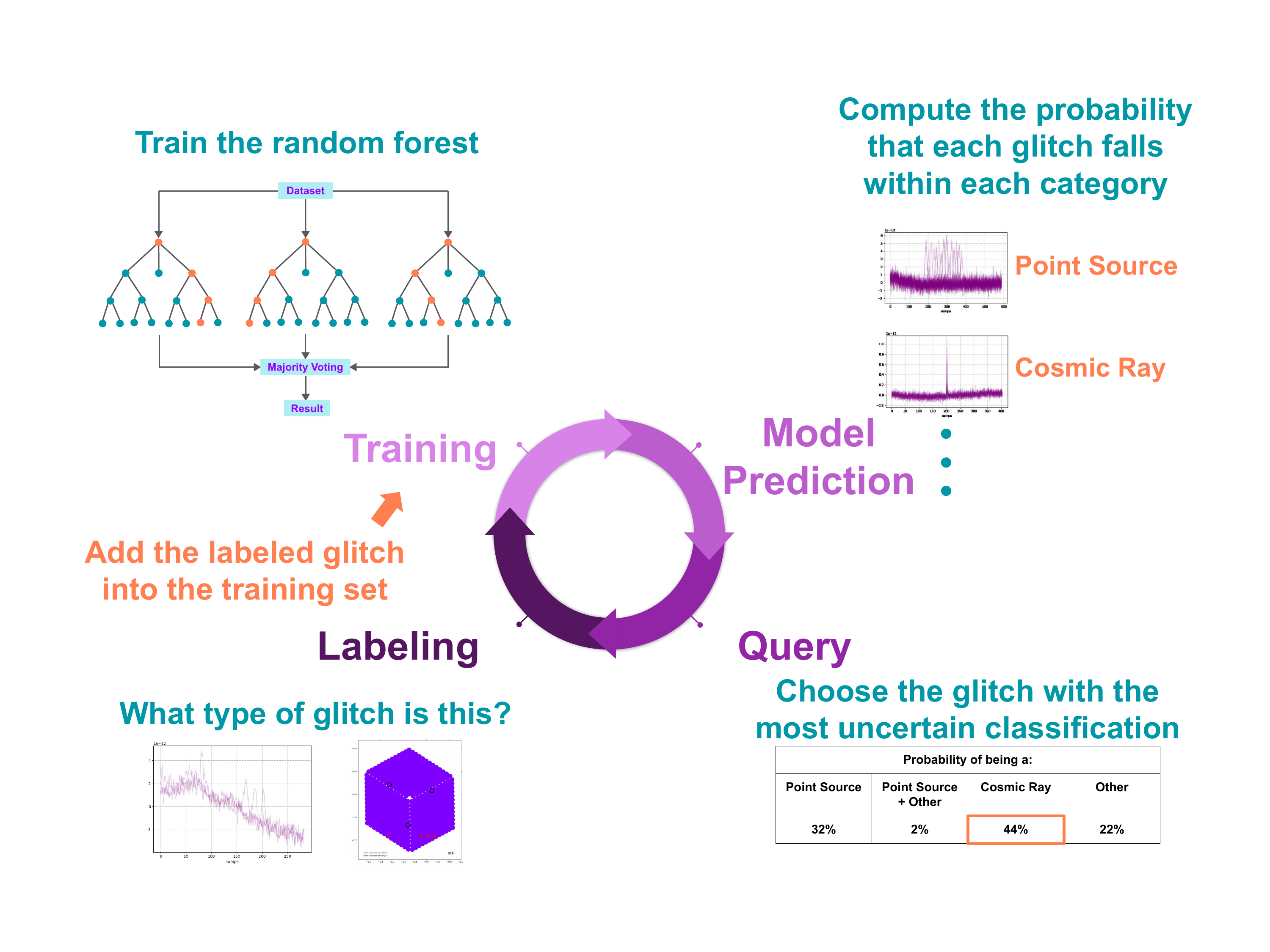}
\caption{Diagram of the AL process. We iteratively first train the forest on a labeled training set, then determine the predictions of glitch types in a separate dataset. For each of the predicted labels in that separate dataset, we determine which glitch label is most uncertain via uncertainty sampling, and then label the glitch (if not already labeled), and add it into the training set. This process continues until we have reached our desired level of classification accuracy.} 
\label{fig: AL diagram}
\end{figure*}

\begin{figure*}
    \centering
    \begin{subfigure}[b]{0.3\textwidth}
   \includegraphics[width=\textwidth]{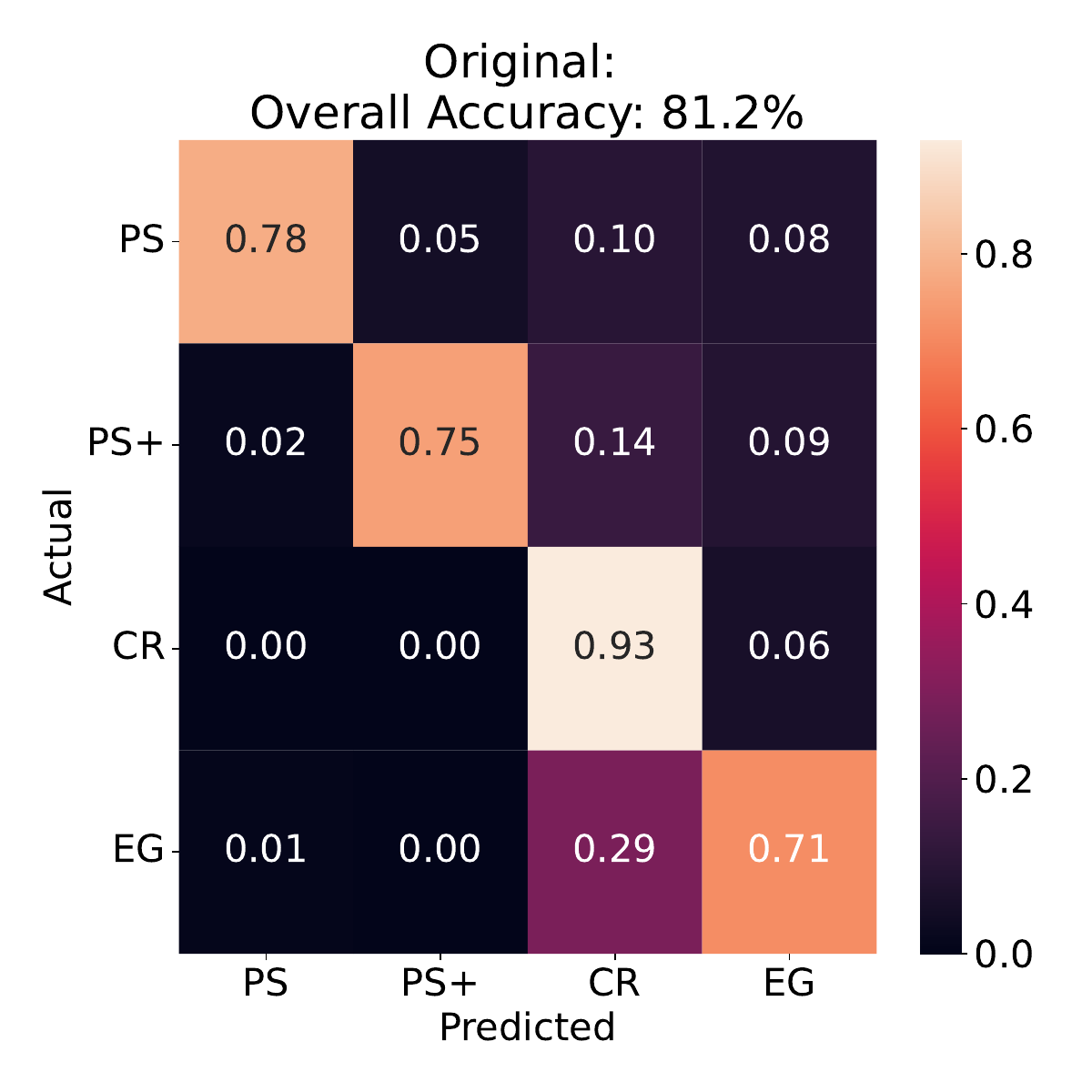}
	\end{subfigure}
 \begin{subfigure}[b]{0.3\textwidth}
   \includegraphics[width=\textwidth]{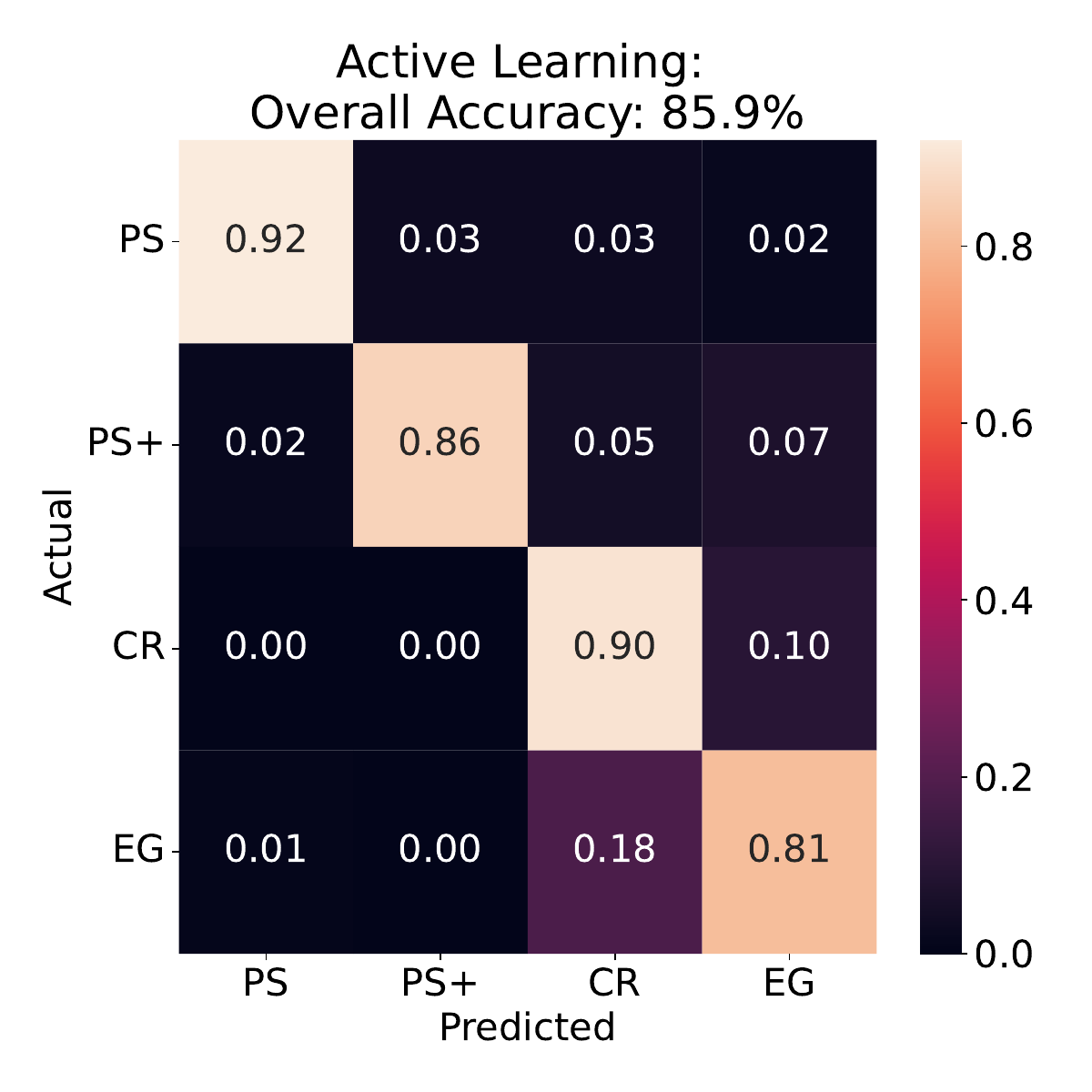}
	\end{subfigure}
 \begin{subfigure}[b]{0.3\textwidth}
   \includegraphics[width=\textwidth]{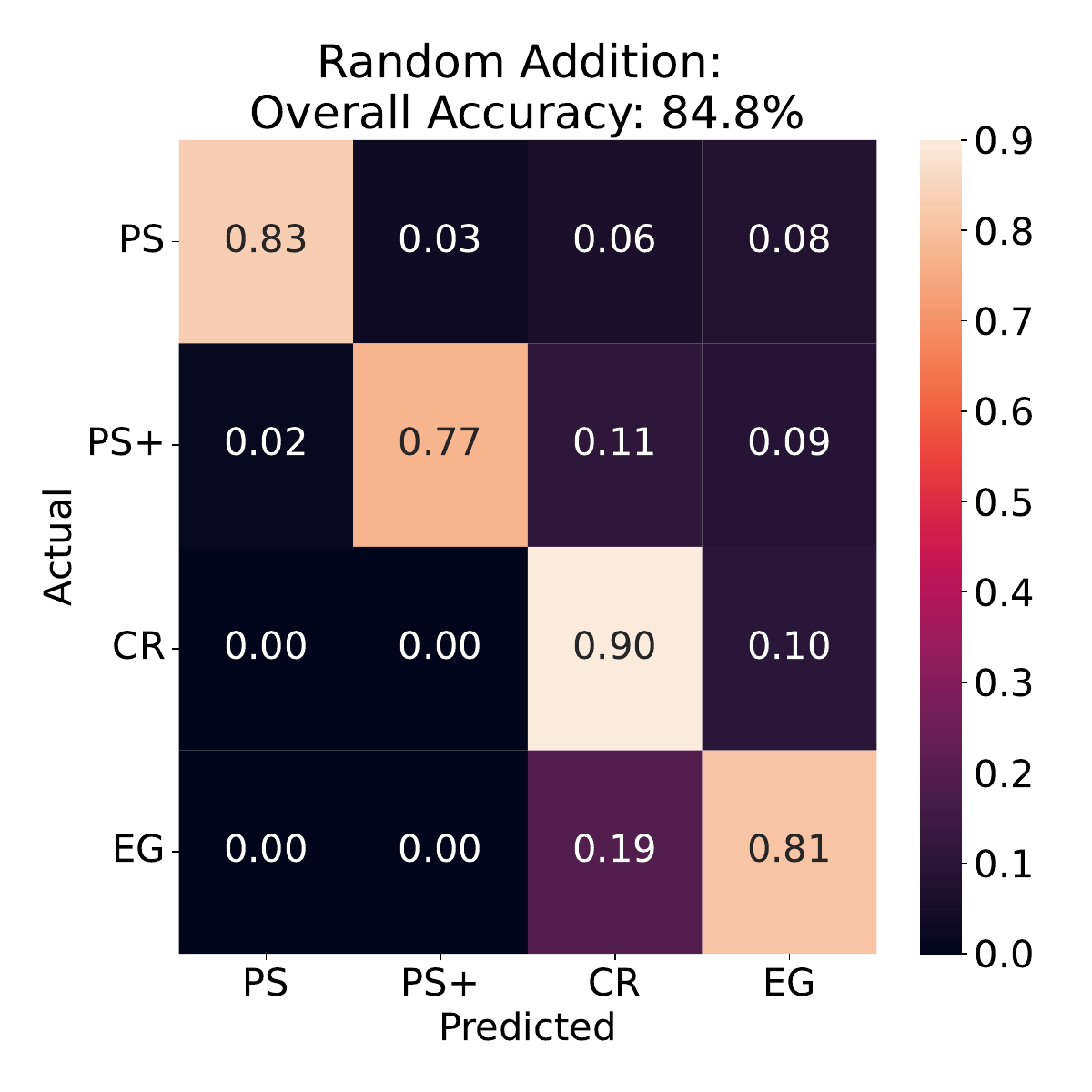}
	\end{subfigure}
    \caption{Exploring the effect of AL on the accuracy of the glitch classifier. Left: confusion matrix using a training set of 2280 points, solely consisting of data in PA5 (90 GHz) and 2019. Additional confusion matrices are shown after adding 300 points to the training from 2021 in PA4 and PA5, as well as both 90 GHz and 150 GHz via AL (middle) and random selection (right). There was an increase of $\sim 5\%$ points in overall accuracy when choosing points via AL, as opposed to $\sim 4\%$ points when they are chosen randomly. Importantly, the PS and PS+ accuracies increase by 9\% points more when choosing points with AL compared to random.}
    \label{fig:AL}
\end{figure*}

Once we have obtained a representative sample for training and testing the random forest, we need to ensure we are robust to changes in the snippets from different years (observation seasons), observational frequency bands, detector array configurations, etc., which requires further refinement of our test/training split. In order to ensure we are robust to these changes, we use an active learning (AL) prescription, in particular the \texttt{modAL} \citep{modAL2018} learning framework. AL is a machine learning approach where unlabeled objects are selected for manual classification/labeling depending on specific criteria of the object (e.g. the classification ambiguity between types in a given set, or an estimate of the Bayesian classification of the object type), as opposed to selecting the point for labeling randomly from the unlabeled set. We employed the \texttt{modAL} package to choose objects using a selection criterion based on which objects are most ambiguous, or in other words, objects for which the maximum classification probabilities are the lowest. In doing so, we can leverage human/manual classifications where our random forest classifier might be most uncertain. This criterion for selecting objects for manual classification is known as ``uncertainty sampling''. Figure~\ref{fig: AL diagram} shows a diagram of the process. 

From the initial random forest classification, each glitch has four probabilities related to the four classification categories: PS, PS+, CR, and EG. The most likely (highest) probabilities for all glitches are ranked, and the uncertainty sampler takes the lowest value of those probabilities. In other words, the uncertainty sampler selects a glitch that the random forest classifier is the least confident in (or most uncertain about) to manually label and therefore add to the AL training set \citep{LEWIS1994148}. This is shown in the bottom right-hand section of Figure~\ref{fig: AL diagram}.

To demonstrate the effects of AL, we tested the random forest on a set of 851 labeled glitches from 2021 at 90 GHz and 150 GHz for detector arrays PA4 and PA5. We begin with a training set composed of 2280 glitches from 2019 at 90 GHz in PA5. We added 300 training points using two methods, AL and then random selection, that were chosen from a set of 2555 glitches from 2021 at 90 GHz and 150 GHz in PA4 and PA5. Note that the 851 and 2555 glitches were chosen by a 25/75 \% random split from an initial data set of 3406 glitches. We chose to add 300 samples as the accuracy increase for the PS and PS+ categories, which we care most about, plateau. Additionally, we want to test realistic settings for the future where ideally we will need to label a small number of additional training points for new seasons of data, additional arrays, etc. We trained three random forests using the three training sets: original, addition with AL, and random addition. The test set is then classified using each of the forests to compare their performance. It is important to note that the test set and additional training points are comprised of data from different arrays, frequencies, and years which could have different noise properties or detector performance than the original training set with only data from 2019 at 90 GHz for PA5. The test set remained the same throughout, in order to ensure that the accuracy was not increasing due to removing difficult-to-classify test points. The class distribution for the test, original training, training points that were available to add, training with AL addition, and training with random addition sets can be seen in Table \ref{tab: class distribution}.\footnote{All of the training and test sets can be found at \url{https://phy-act1.princeton.edu/public/data/dr6_tod_v1/}.}

The results can be seen in Figure~\ref{fig:AL} where the left plot shows the confusion matrix that arises from classifications based on the initial training set. Each value in the confusion matrix indicates the decimal percentage of each glitch category on the $Y$-axis being predicted as the category shown on the $X$-axis. For perfect performance, the resulting confusion matrix would be an identity matrix. The middle and right plots show the confusion matrices where additional training points have been chosen through AL (middle) or random selection (right). In both cases, the overall accuracy of classification increases modestly, by $\sim 5\%$ and $\sim 4\%$ points, respectively. We note that the classification accuracy for the PS category increased significantly, by close to 14\% points, when using an AL adjustment and by only 5\% points when supplementing the training set randomly.

Similarly, the accuracy for PS+ increased by 11\% points when utilizing AL, as opposed to 2\% points with random selection. This is promising: As the volume of data increases with successive observing seasons, new detector arrays, and looking to future telescopes such as SO, we are now motivated to use AL to decrease the amount of manual labeling needed in a training set. AL also allows us to ensure that we are choosing test points that are difficult for the trained random forest to classify, and therefore gaining a better understanding of its performance.

\subsection{Classification Performance Metrics}

\begin{table}
    \centering
    \caption{Class distribution for testing the effects of random addition vs. AL.}
    \label{tab: class distribution}
    \begin{tabular}{|l|c|c|c|c|c|}
    \cline{2-6}
    \multicolumn{1}{l|}{} & \multicolumn{5}{c|}{\textbf{Number of Snippets}}             \\ \hline
        \textbf{Data Set} & \textbf{Total} & \textbf{PS} & \textbf{PS+} & \textbf{CR} & \textbf{EG} \\\hline
        Test & 851 & 63 & 44 & 363 & 381 \\
        Original training & 2280 & 239 & 121 & 1040 & 880 \\
        Available for addition & 2555 & 170 & 156 & 1068 & 1161 \\
        AL training & 2580 & 272 & 167 & 1119 & 1022 \\
        Random training & 2580 & 255 & 138 & 1142 & 1045 \\\hline
    \end{tabular}
\end{table}

To test the performance of the classifier, we also examined the precision, recall, and F-1 score for each of the classification categories. As illustrated in Equation~\ref{eq:prf1}, precision is the proportion of glitches predicted to be within a category that were predicted correctly, while recall measures how many glitches of a certain category were correctly predicted. A high precision score therefore indicates a low false positive rate (few objects incorrectly identified), whereas a high recall indicates a low false negative rate (few objects missed). The F-1 score is a harmonic mean of the precision and recall of the forest. They are given by: 
\begin{equation}
    p = \frac{\text{tp}}{\text{tp + fp}}, \qquad
    r = \frac{\text{tp}}{\text{tp + fn}}, \qquad
    \text{F-1} = \frac{2p\times r}{p + r}, \label{eq:prf1}
\end{equation}

\noindent where ``tp" is the number of true positives, ``fp" is the number of false positives, ``fn" is the number of false negatives, $p$ is the precision, $r$ is the recall, and ``F-1" is the F-1 score.

\subsection{Fine-tuning the Random Forest}

We explored the performance of the random forest as a function of the number of trees and the ``max depth'' (which is the maximum number of times each decision tree is allowed to split) of the forest classifier. The number of decision trees used was varied between 25 and 175 (preliminary tests outside that range showed minimal impact in performance) and the max depth varied within an initial range of 2 to 32. The performance was assessed using the overall accuracy of the classifier, the overall out-of-bag (OOB) score, and F-1 score. The random forest approach uses bootstrap aggregation to train each tree, a process whereby the training set for each decision tree is produced by sampling from the original training set with replacement. This avoids overfitting of the random forest to the training data, which can lead to a model that is does not perform well on new (test) data. The accuracy of each tree is computed with the samples that were not included in its bootstrapped training set. The average accuracy score from all the trees is known as the OOB score and is used to characterize the ability of a method like random forests to make predictions on data that were not included in the training set \citep{OOB}. The performance metrics discussed above plateau and do not improve when the forest reaches a size of 50 trees and a max depth value of 15; therefore, we fixed those hyperparameters of the random forest method to these values.

\subsection{Feature Importance}

When using the various statistics described in Section \ref{sec:classification_algorithm}, it is important to determine which of the features are more instructive in the classification step, to ensure that we understand how the classifier weights them. Low-importance features can potentially be removed from a feature set, and high-importance features can be examined to identify the physics behind why the features lead to more accurate classifications. It also aids in our understanding of misclassified glitches; see Section~\ref{ssec:class results} for more details.

\begin{figure}
\centering
\includegraphics[scale=.3]{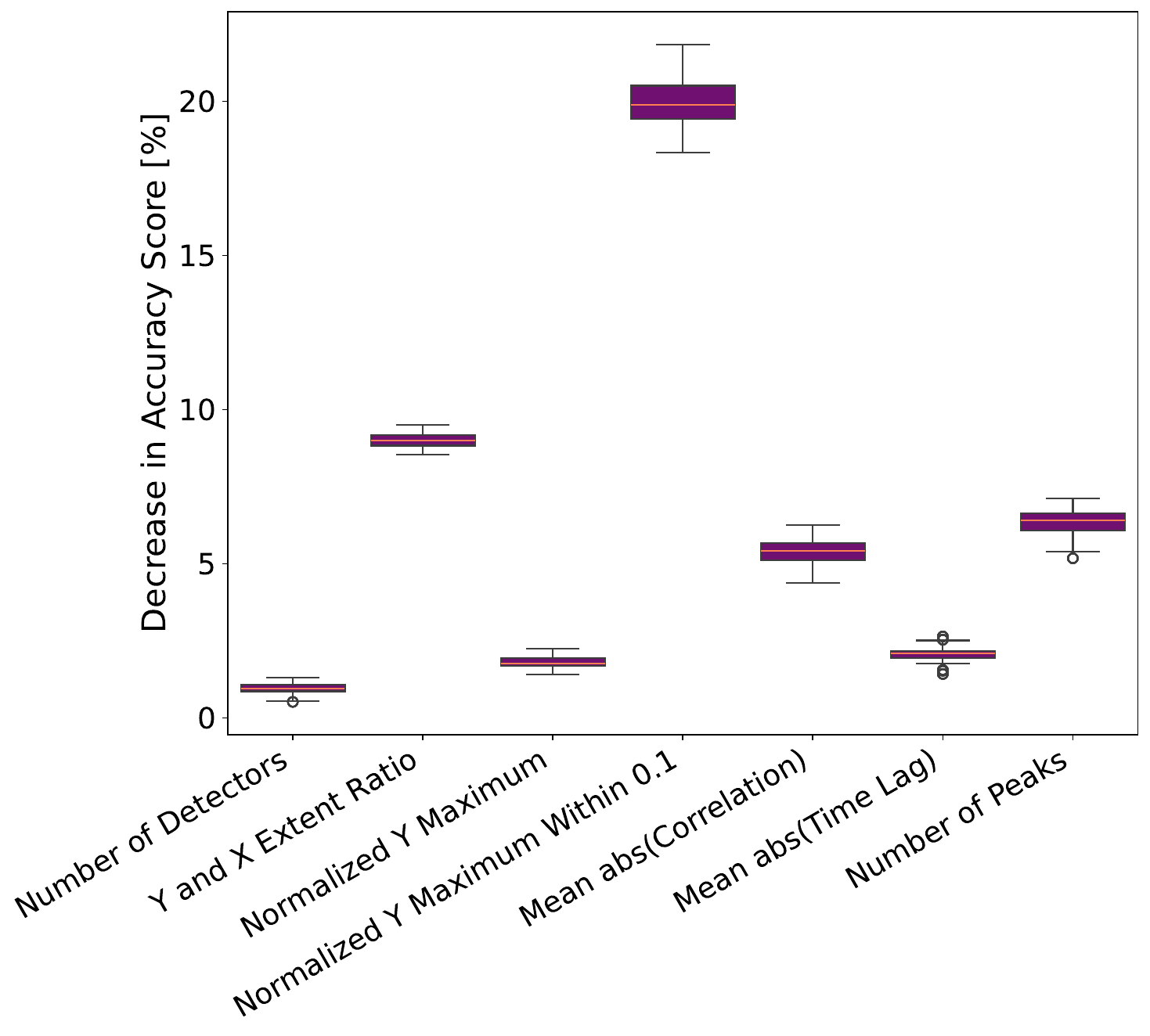}
\caption{The decrease in accuracy score when perturbing each of the summary statistics used for classification, computed via the permutation feature importance method over 50 iterations. The median of these distributions is shown in coral. The two features with the largest impact, and therefore most important, are the $Y$ and $X$ extent ratio and the normalized $Y$ maximum within 0.1$\degree$.}\label{fig:stats importance}
\end{figure}

We compute the relative importance of the various features described above to train our random forest classifier using the ``permutation feature importance'' method \citep{randomforest}. This method randomly shuffles the values of each feature and computes the change in the model performance for those values. The impact of this on the performance of the classifier indicates the importance of each feature in determining the classifications. We randomly shuffle the features 50 times each to reduce the variability of single permutations, and show the distributions of the feature importance for each statistic in Figure~\ref{fig:stats importance}. A larger decrease in the accuracy score indicates a higher importance of the feature, as it shows that changing the values of those statistics would have a more significant impact on the classification results (and vice versa). For our model, the focal plane statistics tend to have the largest impact on the accuracy score. This is discussed further in section \ref{ssec:class results}.

\section{Stellar Flare Simulations}\label{sec:sims}

\begin{figure}
    \centering
    \includegraphics[scale=0.49]{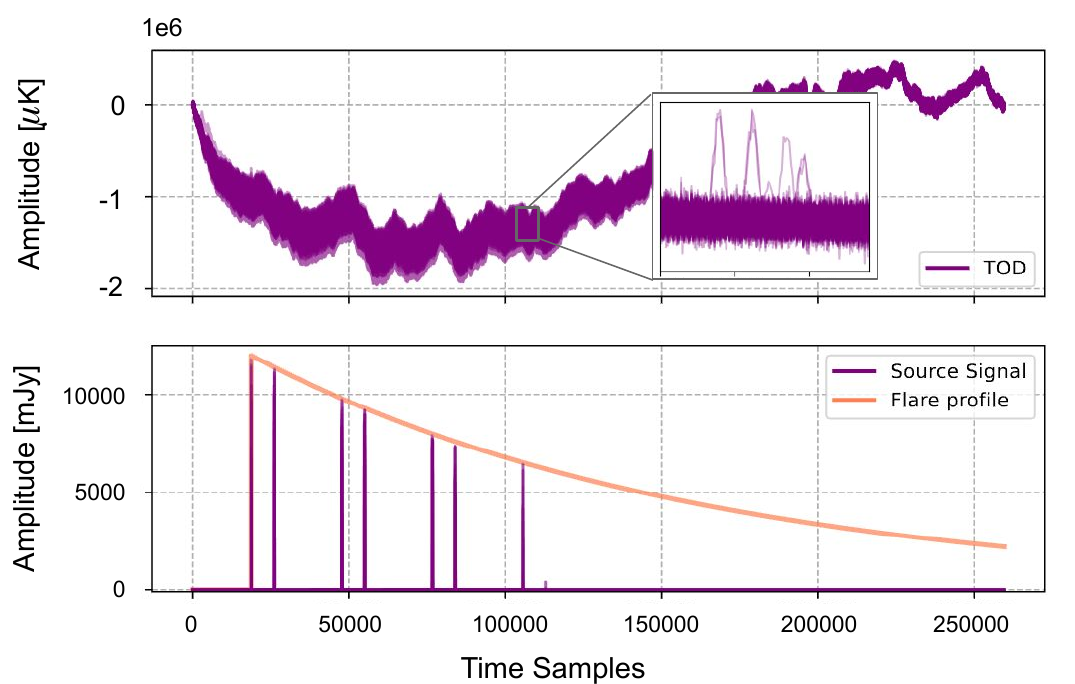}
    \caption{TOD containing an injected simulated stellar flare with an initial amplitude 12,000\,mJy. The half-life of the stellar flare is $h = 250$\,s. The $X$-axis is the time sample (with a sampling rate of 400\,Hz). \textit{Top:} TOD with the injected source. \textit{Bottom:} amplitude profile of the injected flare in coral, with signal that is injected into the TOD in purple.}
    \label{fig:sim_example}
\end{figure}

To test the classification pipeline on a point source with known position and amplitude, we inject simulated signals of stellar flares into real TODs. This allows us to test our classification pipeline on timestreams that contain real noise plus the addition of signals from simulated sources. We then use the known pointing matrix, which describes the sky coordinates of the telescope pointing as a function of time, to project an instrumental beam profile, as is appropriate for a point source, into the TOD timestreams. This process projects the 2D spatial beam into the timestreams for all detectors that scan over the simulated source.

We vary the amplitude of the injected source over time as:
\begin{equation}
\begin{cases}
    S(t) = 0 &, t < t_0 \\ 
    S(t) = A_0 \exp[-\frac{\ln(2)}{h}(t-t_0)] &, t \geq t_0,
\end{cases}
\end{equation}
where $A_0$ is the initial amplitude of the flare, $h$ is the half-life of the decaying flare, and $t_0$ is the start time of the flare, which we select to be when the source enters the focal plane for the first time in the TOD. Stellar flares can display a variety of behaviors across the electromagnetic spectrum and can be fit with different profiles. For instance, \citet{MacGregor} fit two millimeter flares detected by the Atacama Large Millimeter/submillimeter Array \citep{Wootten:2009} with Gaussian profiles. \citet{Salter}, on the other hand, used an exponential curve to fit the decaying millimeter flux of the equal-mass, highly eccentric, close binary system DQ Tau observed with the Institute for Radio Astronomy in the Millimetre Range Plateau de Bure Interferometer \citep{Guilloteau:1992}, the Combined Array for Research in Millimeter-wave Astronomy \citep{Beasley:2003}, and the Submillimeter Array \citep{Ho:2004}. In this paper, we opt for a simple model in which an instantaneous jump in flux is followed by an exponential decrease.

Figure~\ref{fig:sim_example} shows an example of a stellar flare injected into a TOD. The simulations are run for sources with an initial amplitude of 12,000 and $6000$\,mJy and flare profiles with half-lives of $h=750, 500, 250, 100, 25, 5, 0.5$ seconds. These amplitudes were chosen as this specific TOD has an average noise level of $\sim 700$\,mJy, and we wanted to test a case where most detectors would be well above the $\mathrm{SNR} > 10$ cut-off and a case closer to the $\mathrm{SNR}\sim 10$ limit. These half-lives were chosen to cover a wide range of flare durations. The shortest ($0.5$\,s) half-life reproduces a situation where the flare decays significantly within a single scan, whereas the longest ($750$\,s) half-life was chosen to maximize the number of detections within the 11-minute TOD. The TODs are loaded in raw data acquisition units then converted into mJy, at which point the source signal is injected. The TOD is then converted to $\upmu$K to be compatible with TOD loading and cuts packages, at which point it can be run through our pipeline.

\section{Results and Discussion}\label{sec:results}

\subsection{Classification Results} \label{ssec:class results}
 
\begin{table*}
    \centering
    \caption{Precision, recall, and F-1 score values for all glitch categories using 50 trees and a max depth of 15 on a set of 3865 labeled glitches, where 760 are obtained from the 2019 data and 3105 from 2021 observations.}
    \label{tab: performance}
    \begin{tabular}{|l|c|c|c|c|}
    \hline
        \textbf{Glitch Category} & \textbf{Number of Snippets} & \textbf{Precision [\%]} & \textbf{Recall [\%]} & \textbf{F-1 Score [\%]} \\\hline
        PS & 282 & 96.0 & 93.6 & 94.8 \\
        PS+ & 207 & 96.6 & 82.6 & 89.1 \\
        CR & 1672 & 87.2 & 92.6 & 89.8 \\
        EG & 1704 & 91.3 & 87.7 & 89.4 \\\hline
    \end{tabular}
\end{table*}
  
\begin{figure}
\centering
\includegraphics[scale=.35]{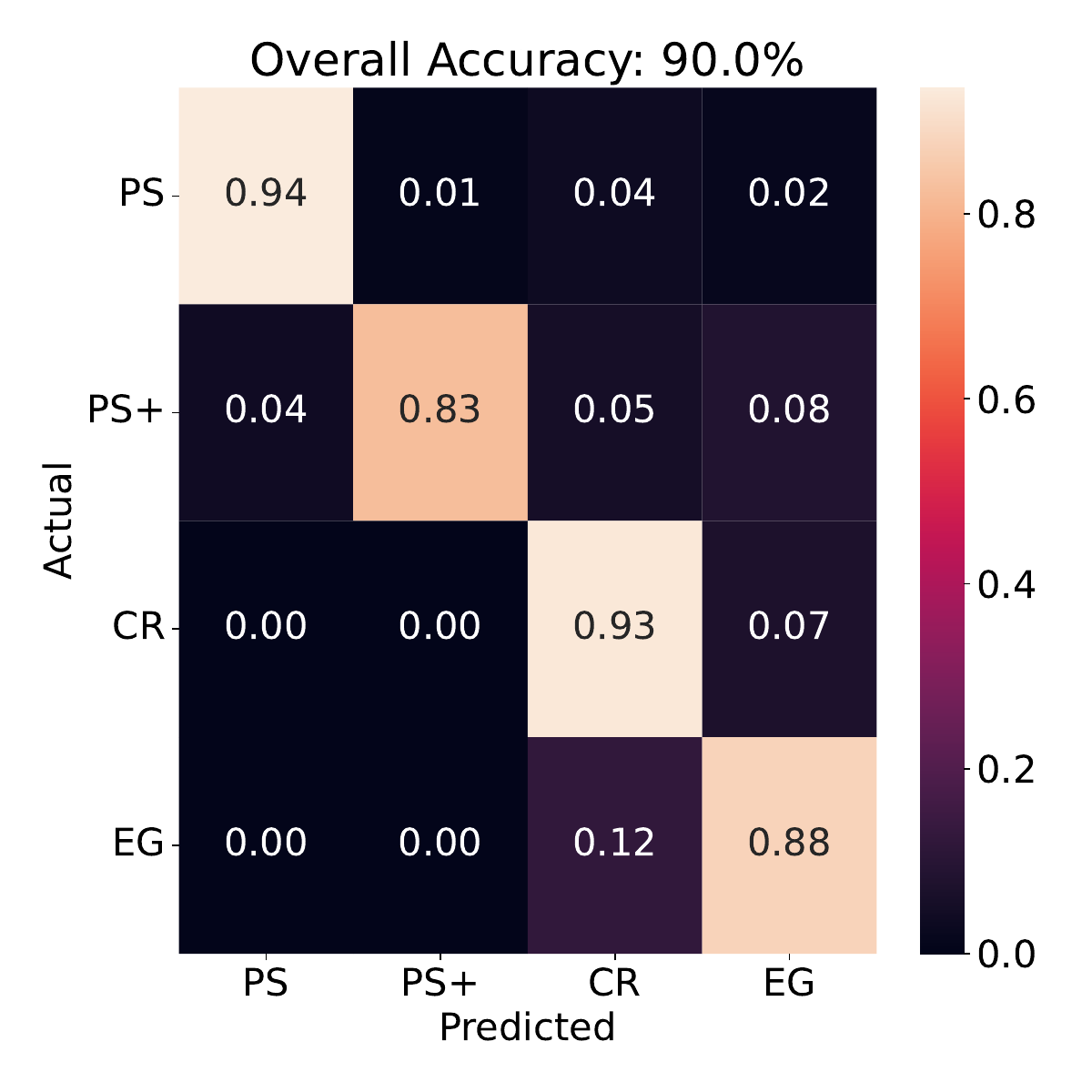}
\caption{The confusion matrix from the random forest classification on a set of 3865 labeled glitches, 760 from 2019 and 3105 from 2021. Note that each decimal percent is rounded to two decimal points.} \label{fig: confusion matrix}
\end{figure}

We tested our trained forest on a set of real data to determine how effective it is at classifying true snippets, all labeled as outlined in Section~\ref{ssec: labeling}. Our training set includes 2580 glitches which consist of 272 PS, 167 PS+, 1119 CR and 1022 EG objects respectively. Of these glitches, 2280 are from 2019, 90\,GHz from PA5, and the remaining 300 are from 2021, 90\,GHz and 150\,GHz from PA4 and PA5. For testing, we had 3865 glitches: 282 PS, 207 PS+, 1672 CR, and 1704 EG. This consisted of 760 glitches from 2019, 90 GHz, and PA5 as well as 3105 from 2021, 90\,GHz and 150\,GHz from PA4 and PA5. Figure~\ref{fig: confusion matrix} shows the resulting confusion matrix. Our overall accuracy is 90\%, and notably, we achieved a 94\% accuracy for PS objects. For CR objects, $\sim7\%$ were classified as EG and $\sim12\%$ of the EG objects were classified as CR. This is expected as some electronic glitches, such as large-area heat-ups of the focal plane, look similar to cosmic rays and vice versa.

The performance metrics for our test set can be seen in Table~\ref{tab: performance}. All the F-1 scores across the four categories are greater than 89\%, which indicates a well-performing model. We consider a well-performing model to be one where the metrics discussed above are $\geq 80\%$; however, we also favor as high a value of precision as possible, as false detections of point sources that are left in the TODs will have a negative impact on the maps.
 
\begin{figure*}
    \centering
  \begin{subfigure}[b]{0.3\textwidth}
   \includegraphics[width=\textwidth]{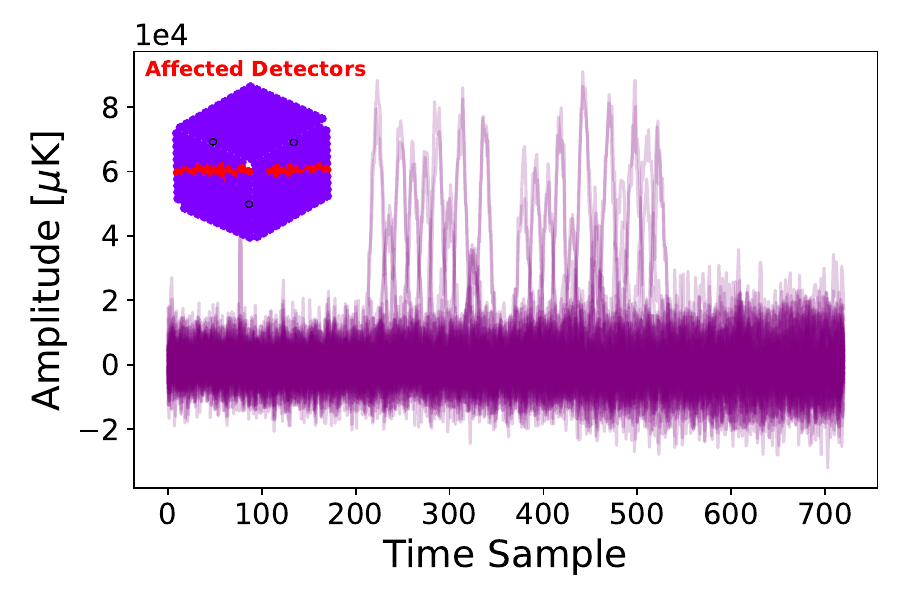}
	\end{subfigure}
  \begin{subfigure}[b]{0.3\textwidth}
   \includegraphics[width=\textwidth]{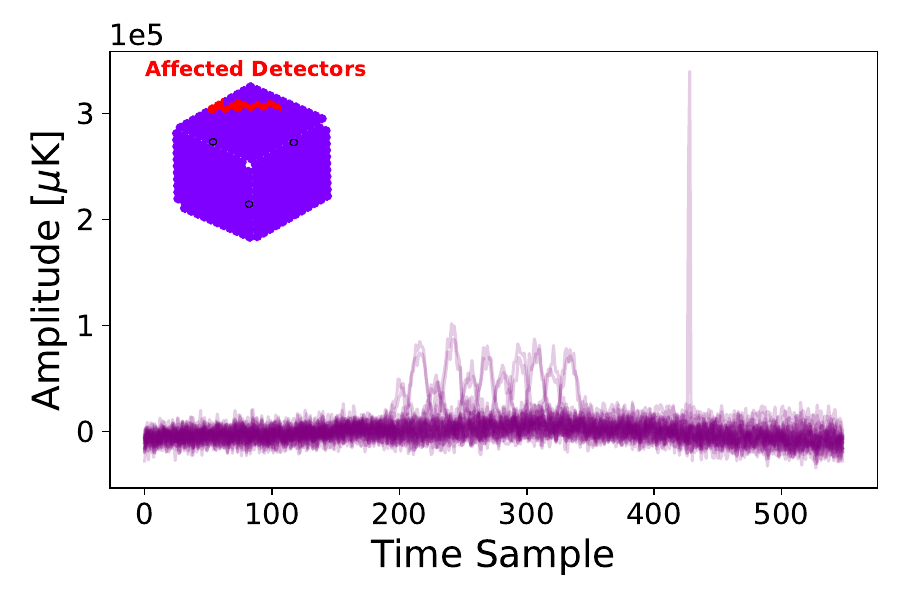}
	\end{subfigure}
 \begin{subfigure}[b]{0.3\textwidth}
   \includegraphics[width=\textwidth]{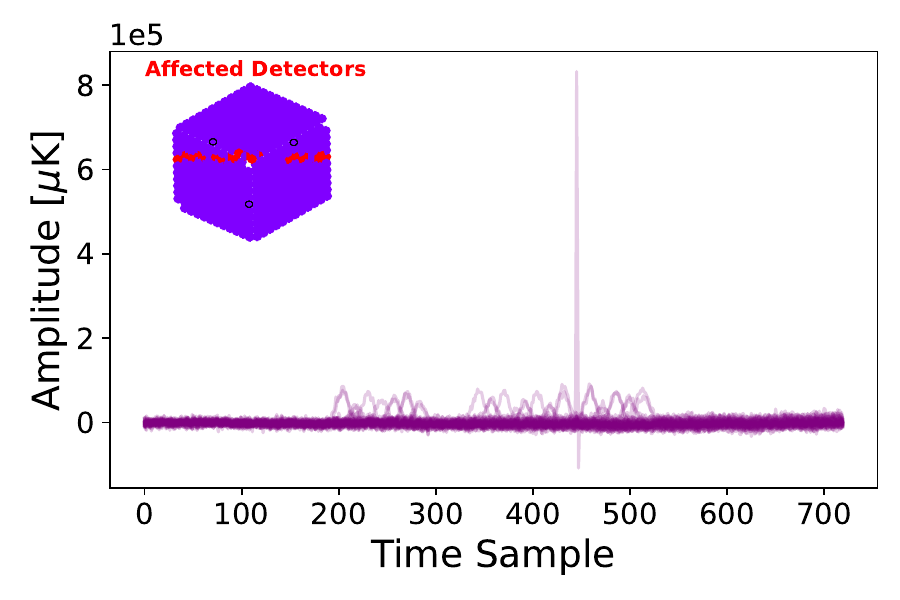}
	\end{subfigure}
 \begin{subfigure}[b]{0.3\textwidth}
   \includegraphics[width=\textwidth]{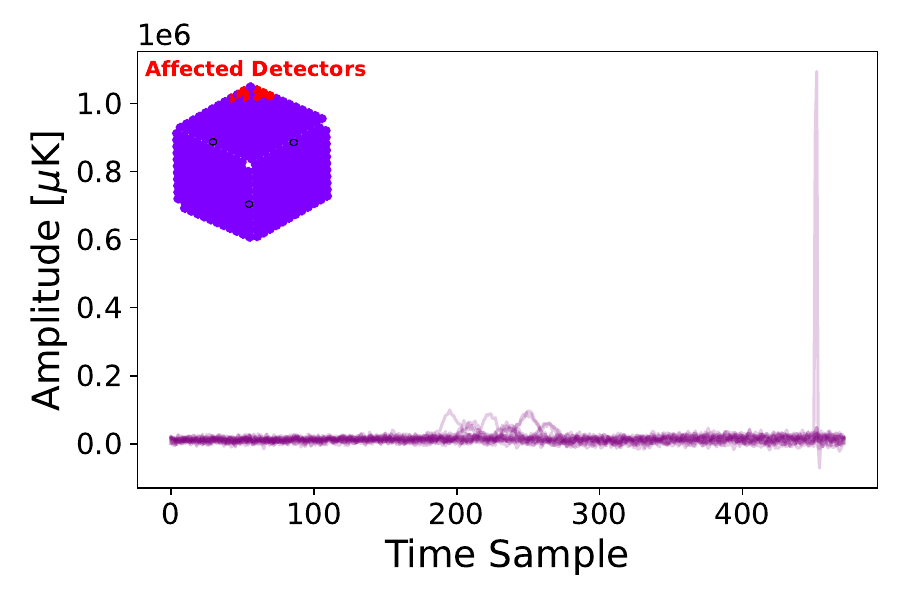}
	\end{subfigure}
 \begin{subfigure}[b]{0.3\textwidth}
   \includegraphics[width=\textwidth]{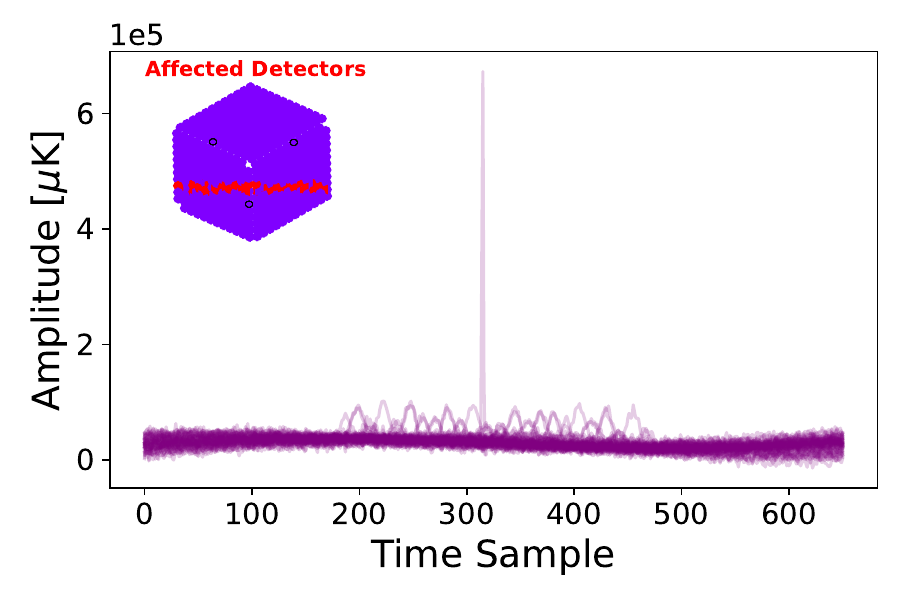}
	\end{subfigure} 
 \begin{subfigure}[b]{0.3\textwidth}
   \includegraphics[width=\textwidth]{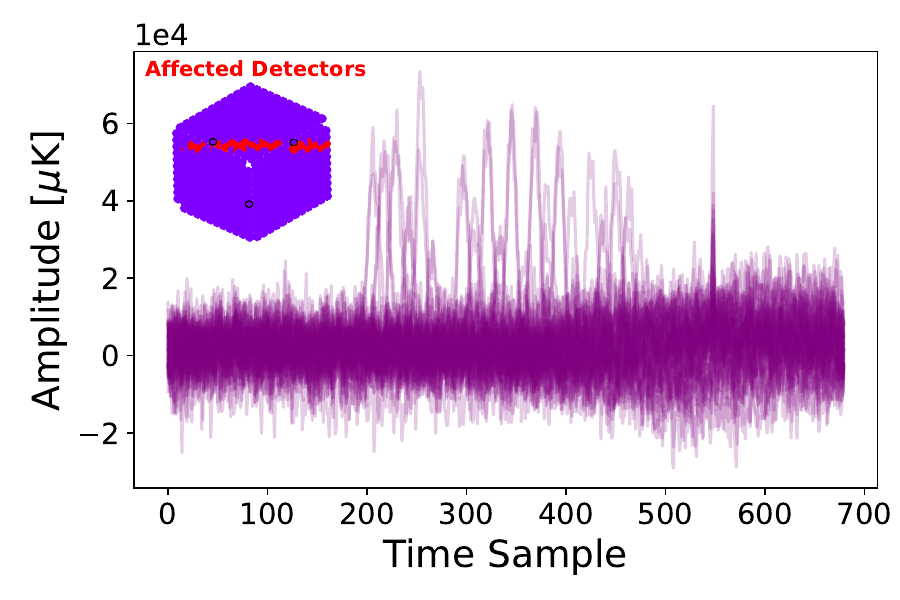}
	\end{subfigure}
    \caption{All glitches that were incorrectly classified as a PS with a probability greater than 70\%. The TODs for the affected detectors are shown in purple, and each contains a focal plane image showing the affected detectors in red. Notice that these are all PS+ occurring where the additional glitch is not very distinct from the PS on the focal plane.}
    \label{fig:incorr ps o}
\end{figure*}

We consider a classification of an individual object with a true underlying type to be ``high-probability'' if it is predicted to be a particular type with a probability, defined as the percentage of decision trees in the forest that predict the glitch to be a given type, to be greater than 70\%. After filtering for high-probability classifications in our test set, we are left with six PS+ objects that were mistakenly classified as PS objects, and examine these individually to understand the misclassifications. The TODs and focal plane positions for these snippets are shown in Figure~\ref{fig:incorr ps o}. In these TODs, another glitch appears alongside a point source, but the detectors affected by the other glitch are located near to the PS in the focal plane. This resulted in the normalized $Y$ maximum within 0.1$\degree$ statistic for all cases to be equal to one. Similarly, the normalized $Y$ maximum values for these objects range from 0.27 to 0.51 and the $Y$ and $X$ extent ratios range from 0.06 to 0.24, which is consistent with values observed from objects with the PS designation rather than PS+. As described in Section~\ref{sec:classification_algorithm}, the three statistics mentioned above are usually the only ones that separate cleanly between the PS and the PS+ categories, while the other statistics used for classification are consistent with expected values for both types. In this case, the fact that the statistics we input as the features of the random forest suggest a PS categorization (as is shown in Figure~\ref{fig: dists}) is the dominant reason for the misclassification. We repeat that these three statistics are important for the classification results, as can be seen in Figure~\ref{fig:stats importance}. Future work should be done to determine another statistic to aid classification in these instances. One possibility would be a statistic that identifies spikes in the signal that are beam-shaped as we would expect from a source but not from a cosmic ray or electronic glitch. Note that human bias also affects the results, as our training and testing labels are from human labeling.

\subsection{Improved Transient Search Maps}

In addition to removing glitches from the TODs prior to map-making, our classification pipeline is also able to classify astrophysically interesting bright sources from the TODs and generate a list of detection times and locations of these sources. This represents a complementary method to map-based transient search algorithms \citep[e.g., those presented in][]{Biermann:2024emf}. For map-based transient searches, ``depth-1'' maps, or maps from a single array and frequency in which each point in the map drifts only once through the focal plane---roughly speaking, a single observation session---are used. \citealt{Biermann:2024emf} provides details on the ACT depth-1 maps and the results of blind transient searches on these maps. Current map-based searches have focused on finding faint sources outside the Galactic plane. This is partially due to the limitations of identifying many sources that are spatially close together in the maps, as occurs within the Galactic plane. Because our classification pipeline processes each TOD snippet individually, we do not, however, encounter this limitation to the same extent as map-based searches. 

\begin{figure}
\centering
\includegraphics[scale=.25]{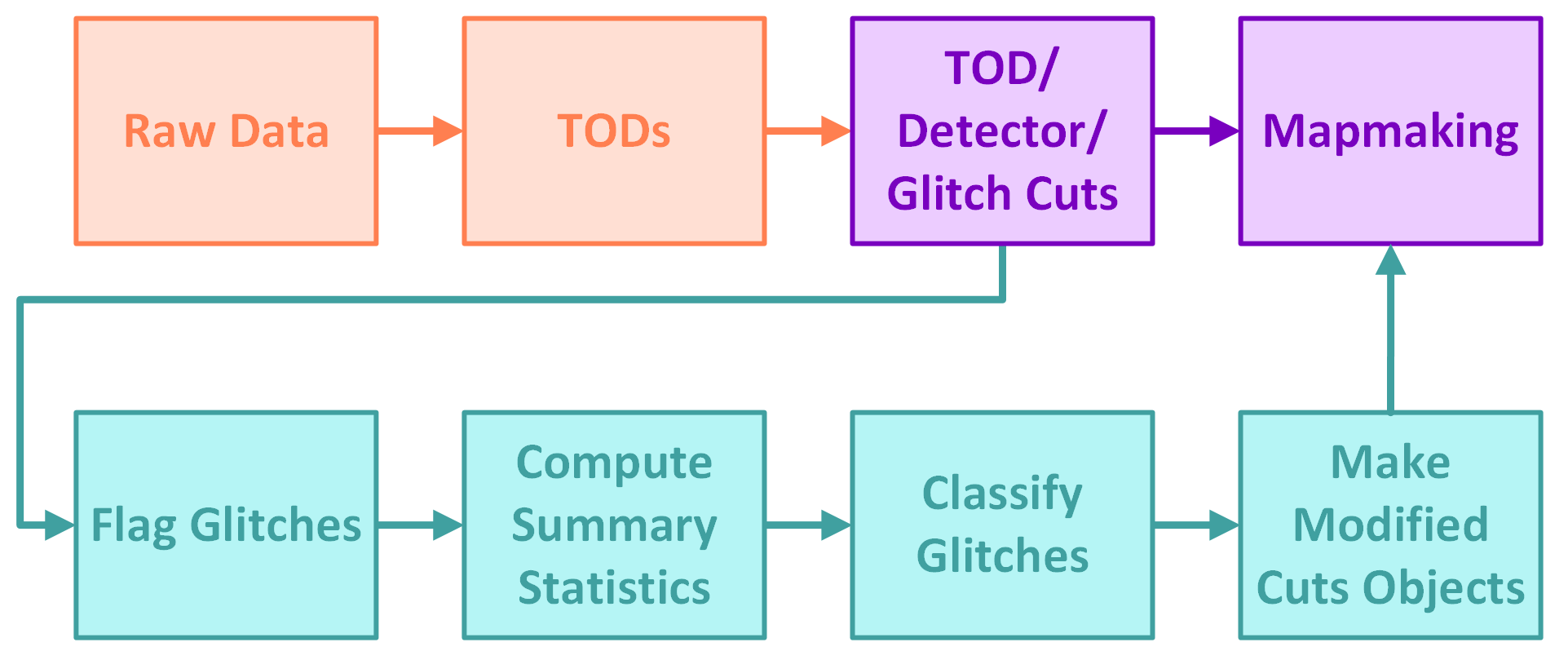}
\caption{Flow chart showing how the ACT pipeline goes from raw data to maps. The portions we have modified are shown in purple, and the new glitch classification pipeline is shown in teal.} \label{fig: flow chart}
\end{figure}

\begin{figure}
\centering
\includegraphics[scale=.4]{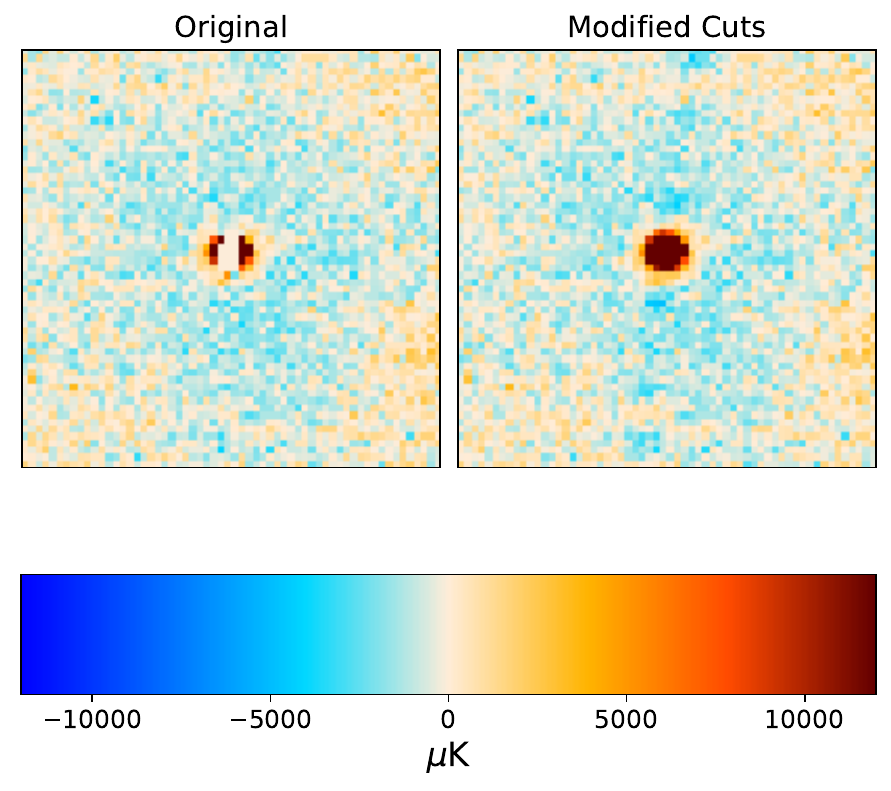}
\caption{Half-degree cut-outs of ACT depth-1 maps with a simulated source with an amplitude of 12,000\,mJy and a 5000\,s half-life. \textit{Left column:} the original depth-1 map made with ACT standard cuts algorithm. \textit{Right column:} the same depth-1 map but with features identified by the random forest to be a PS with high-probability ($>70$\%) included. Note that the source was almost completely cut out using the original method, which could result in it not being detected during map-based searches, but using our modified pipeline, it was recovered.} \label{fig: amp 12000 only}
\end{figure}

To demonstrate that our classification pipeline can prevent genuine sources from being removed from depth-1 maps, we modify the ACT cuts software to not remove any source which has a probability of being a PS greater than 70\%, as determined by the random forest, and compare the results to what is obtained with the original cuts software, doing so on a series of stellar flare simulations described in Section~\ref{sec:sims}. The steps of this modified pipeline can be seen in Figure~\ref{fig: flow chart}, where the existing portions of the pipeline that we modified are shown in purple and our new glitch classification pipeline is shown in teal. In order to demonstrate the efficacy of our method, we generate a bright, relatively stable source that stays at $\mathrm{SNR} \geq10$ by injecting a stellar flare with a peak amplitude of 12,000~mJy and a half-life of 5000~s. The results can be seen in Figure~\ref{fig: amp 12000 only}. We note that the original pipeline removes the source from the map, whereas our modified pipeline restores the source. A ringing pattern along the scan direction through the source is visible in this map cut-out. This is a form of subpixel model error bias (related to regression dilution) caused by treating the sky as being stepwise constant in each pixel despite the true signal changing smoothly, and is unrelated to our cuts implementation \citep{SigurdRinging}. Additionally, as we are only producing maps for the purposes of source detection, there could be a lack of convergence in the maps which could also produce some of this affect.
 
\begin{figure*}
\centering
  \begin{subfigure}[b]{0.69\paperwidth}
   \includegraphics[width=\textwidth]{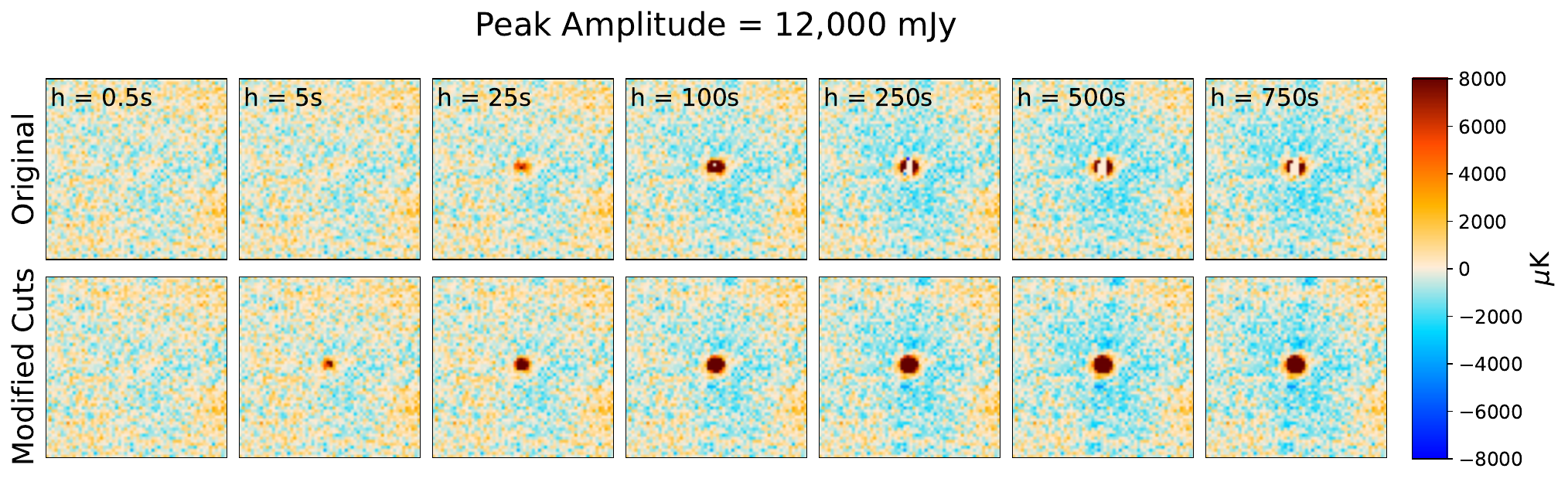}
	\end{subfigure}
  \begin{subfigure}[b]{0.69\paperwidth}
   \includegraphics[width=\textwidth]{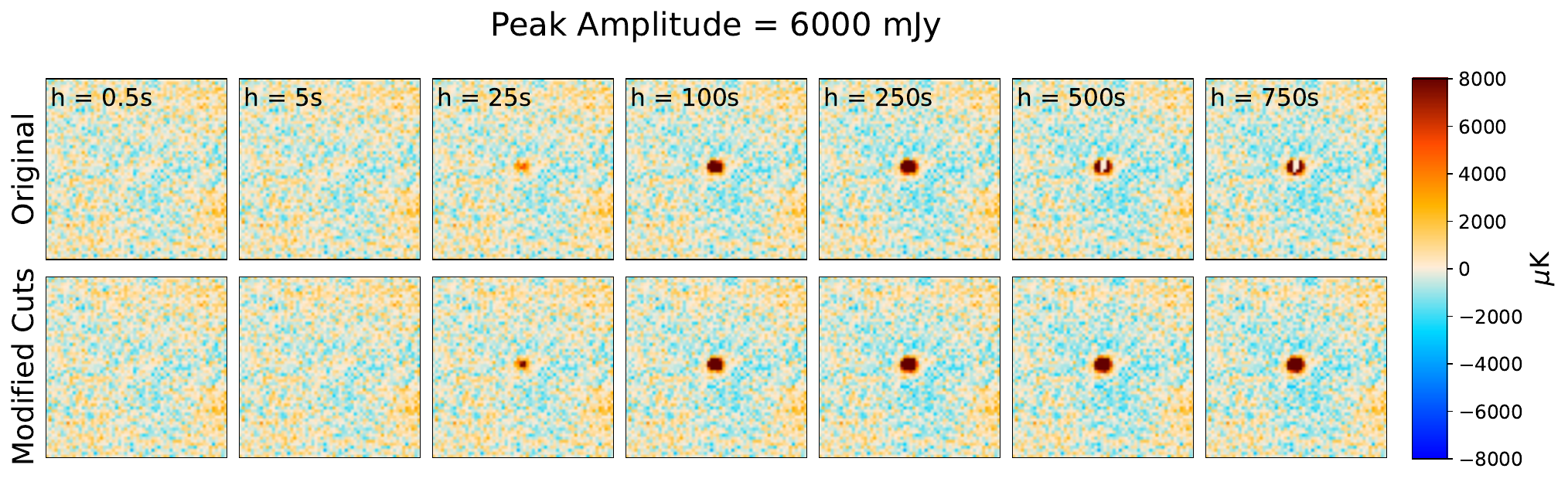}
	\end{subfigure}
\caption{Half-degree thumbnails of depth-1 maps around the simulated stellar flare with peak amplitudes of 12,000\,mJy and 6000\,mJy and half-lives of 0.5\,s, 5\,s, 25\,s, 100\,s, 250\,s, 500\,s, and 750\,s. The original map using the traditional ACT cuts and map-making pipeline is shown on the top row, and the map made with our modified pipeline on the bottom row, per amplitude. \textit{12,000\,mJy:} Notice that the source is not detected in either map for $h = 0.5$\,s; we do in fact detect the source in the TODs but only with a probability of 70\% so it is still removed from the maps. The source has been restored for all other half-lives. \textit{6000\,mJy:} For $h = 0.5$ s and 5\,s we do detect the source but only at a probability of 68\%, so is not detected in either map. The source has been restored for all other half-lives.} \label{fig: half-life cutouts}
\end{figure*}

To test the ability of our pipeline to identify sources with different properties, we vary the initial peak amplitude and half-life of the simulation. Recall that the cuts algorithm identifies glitches by locating spikes in the TOD with an $\mathrm{SNR} \geq10$; the TOD used in this simulation has an average noise of $\sim 700$~mJy, though individual detectors can have more or less noise than this value. We thus start our exploration with an amplitude of 12,000\,mJy to ensure that we are comfortably above the $\mathrm{SNR}$ threshold and 6,000\,mJy to have a fair number of detectors around the SNR threshold. The top and bottom rows per amplitude of Figure~\ref{fig: half-life cutouts} show maps with each of the original and modified cuts pipelines, respectively. For the 12,000~mJy amplitude source, the source was at least partially cut out of the maps using the original pipeline, but completely restored with the new pipeline for all simulated source half-lives except 0.5\,s, which was classified as a PS, but only with a probability of 70\% (which is at the threshold of what we consider a high-probability detection). In the case of the 6000\,mJy amplitude source, it is above the noise value for most, but not all, of the detectors. In this case, the source is classified as a PS for $h = 0.5$ and 5\,s but only with a probability of 68\%, which is below our detection threshold. The source appears in the map made with the original pipeline and also in our modified pipeline for $h = 100$ and 250\,s and the center of the source is fully restored for $h = 500$ and 750\,s which is necessary for map-based analysis of the source properties. The reason the source can appear in the original maps even though some detectors are cut is because a portion of the detectors see the source below the $\mathrm{SNR}$ threshold, while the ACT cuts algorithm requires that the source has an $\mathrm{SNR} > 10$ for a detector to be cut. There is some utility, however, to being able to identify a source that is only partially detected/cut, as one can measure the amplitude variations between snippets to aid classification of the source when it is later detected in a map-based search. Partial identification of sources also generates a list of times and locations of observed sources that are useful for the map-based search methods described above. Note that we chose the probability threshold in order to reduce the number of false positives for map-making, but for transients searches one could lower this threshold and the source would have been recovered for both amplitudes with a half-life of 0.5\,s.

Finally, we tested our pipeline for an amplitude of 3000\,mJy, which is similar to the amplitude of the brightest transient found in ACT ($2307\pm222$\,mJy; \citealt{Biermann:2024emf,li/etal:2023}). We note, however, that previous transient searches have excluded the galactic plane, where we have found many sources with higher amplitudes that we have successfully identified and restored with the new classification pipeline. The injected source was only identified for $h = 500$ and 750\,s with all other values of half-life resulting in the source remaining below the $\mathrm{SNR}$ detection threshold. We would like to understand if this failure to correctly classify the injected source is due to a limitation of our classification performance or a result of the earlier step in the process where the snippet is flagged by the cuts algorithm. 

To do so we ran the simulated sources through a modified cuts algorithm that we call the ``ideal cuts'' algorithm, that flags every detector that observes the source with an amplitude greater than 500\,mJy, rather than the usual approach used for ACT analysis which requires the signal to be above the $\mathrm{SNR}$ threshold. Given that this is a simulation and hence we know a source is present, we can isolate the ``detection'' versus the ``classification'' parts of our analysis, removing any uncertainty in the detector cuts pipeline and focusing only on the performance of the classification pipeline presented in this paper. Employing this procedure, we were able for each simulated amplitude to detect the injected sources for all values of the half-lives and initial amplitudes we simulated. We provide more details on this test in Appendix Section \ref{sec:ideal cuts}. The failure to detect the source by the detector cuts algorithm is due to the source having an $\mathrm{SNR}$ below the threshold. Due to the importance of summary statistics related to the focal plane statistics in our classification pipeline (as is shown in Figure~\ref{fig:stats importance}), we are able to classify the sources, provided they are detected by our ``ideal cuts'' algorithm even though the source signal is very faint in the TODs. The $\mathrm{SNR}$ cutoff is a known limitation of the cuts algorithm, highlighting the need for future work to create an improved cuts algorithm that would enable our classification pipeline to find lower-amplitude sources. An improved cuts algorithm would enable TOD based classification pipelines, such as the one outlined in this paper, to identify dim sources prior to map-making. This would not only aid in quick identification of possibly interesting sources, but also provide a list of candidate positions and times to search for sources in the depth-1 maps.

\subsection{Point Source Variability}

\begin{figure*}
    \centering
    \includegraphics[scale=0.4]{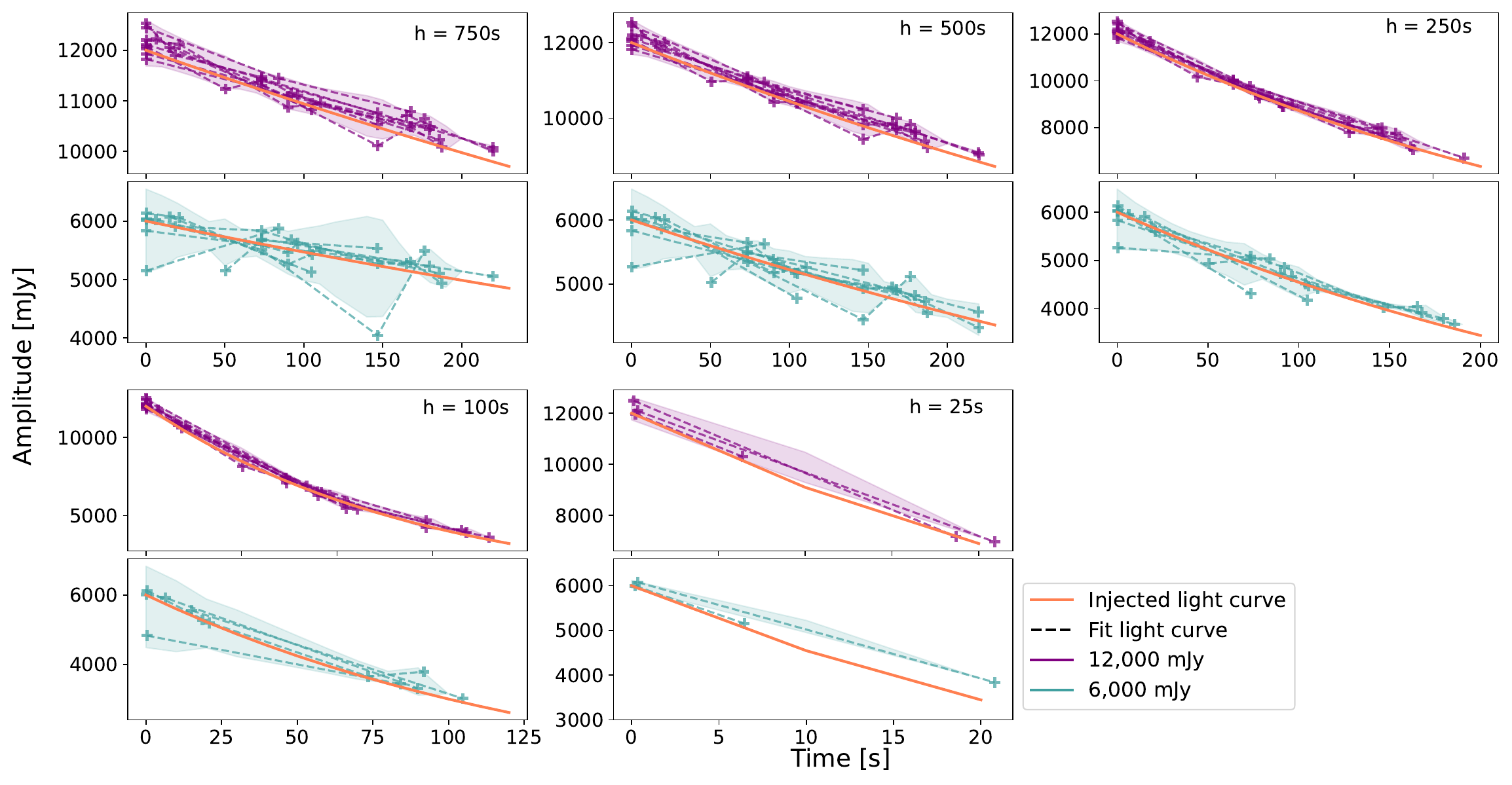}
    \caption{Plot of recovered fit light curves from stellar flare simulations (shown as dashed lines) and the injected flare light curves (coral) for an initial amplitude of 12,000\,mJy (purple) and 6000\,mJy (teal) and half-lives of 750\,s, 500\,s, 250\,s, 100\,s, and 25\,s. The shaded area represents the 2$\sigma$ standard deviation of the recovered fit light curves.}
    \label{fig:all_LCs}
\end{figure*}
  
If a source is observed multiple times, as the focal plane scans back and forth in azimuth at constant elevation, we can measure changes in the amplitude of the source on scales of less than a minute for an ACT-like experiment. This has important implications for transient science, where some sources, like flaring stars, can vary on time scales of minutes or seconds, whereas other sources, such as extragalactic transients like GRBs, TDEs and SNs, vary on much longer time scales of days or more. To issue timely and accurate alerts of the detection of such events to the scientific community, we need our code to be able to measure changes in amplitude on similar timescales. Here, we demonstrate the capabilities of our pipeline to retrieve variability information on classified point sources using the stellar flare simulations described in Section~\ref{sec:sims}.

We model the amplitude of a source in our timestreams with a non-linear least-squares fit to the following model
\begin{equation}\label{eq:amplitude}
    A(\bm{\theta}_{\mathrm{det}}) = A_0 B(|\bm{\theta}_{\mathrm{det}}-\bm{\theta}_{\mathrm{src}}|) + C,
\end{equation} where $A$ is the amplitude measured by a detector observing the sky at R.A./decl position $\bm{\theta}_{\mathrm{det}}$, $A_0$ is the intrinsic amplitude of the source located at position $\bm{\theta}_{\mathrm{src}}$, $B(\theta)$ is the azimuthally-averaged, normalized beam profile of the telescope, and $C$ is an offset to account for the baseline of the signal being nonzero. The independent variable in the fit is $\bm{\theta}_{\mathrm{det}}$ and the fit parameters are $A_0$, $\bm{\theta}_{\mathrm{src}}$ and $C$.
Before performing the fit, we create a copy of the snippet with the peaks excised, and we subtract this copy from the original snippet, leaving only the peaks generated by the point source in the data. This removes any remaining large-scale drift across the length of the snippet, which we find significantly improves the fit. We then discard half of the 0.5\,s buffer from each of the beginning and end of the snippet, i.e., the first and last 100 samples to reduce the amount of noise that goes into the fit, as we know that the snippets have a buffer of 200 samples on either side of the source signal. Finally, we obtain initial guesses for the non-linear fit by averaging the peak amplitude values of all detectors, for the initial value of $A_0$, and the detector positions at their peak values, for $\bm{\theta}_{\mathrm{src}}$; $C$ is initially set to zero. We then run the \texttt{curve\_fit} non-linear fitter as implemented by \texttt{SciPy}.

We group snippets that are classified as high-probability point sources within 0.01$\degree$ of each other (corresponding to the standard error on our point source localization in the timestreams), treating them as the same astronomical object, and then we do the fit described above for each scan of the array across the source. This produces a light curve with one point per scan. Our best-fit light curves to simulated stellar flares injected into real data are shown in Figure~\ref{fig:all_LCs}. The position of the source relative to the scan pattern differs for each TOD, causing the time intervals between the source showing up in successive scans to vary from TOD to TOD. This can be seen in the different segment lengths of the individual best-fit light curves (dashed lines) in Figure~\ref{fig:all_LCs}. For this reason, we linearly interpolated the best-fit light curves into 10\,s time bins. The amplitudes were then averaged in each time bin to obtain a mean and standard deviation, shown with the shaded region in Figure \ref{fig:all_LCs}. We only simulate stellar flares with half-lives above 25\,s because shorter flares decay too quickly to be detected for more than one scan.

To evaluate the level of confidence with which we can detect short-term variations with this method, we calculate the likelihood ratio (or $\Delta\chi^2$, assuming a Gaussian likelihood) of our best-fit light curve from a single TOD to a flat line at the average flux value; the latter is the null hypothesis of no variation in the source. The likelihood ratio can be converted to a p-value for the null hypothesis, or equivalently a significance of the measurement of variability.\footnote{The conversion of the likelihood ratio to a $p$-value assumes that the former has a $\chi^2$ distribution, which is true for numerous samples; in our case the number of samples is low, so the values we report should be taken as approximate.} For uncertainties in our measurement, we use the standard deviation from fits to 11 different TODs into which the same simulated flare is injected. The results of this analysis are presented in Table \ref{tab:sigmas}.
 
\begin{table}[]
\begin{tabular}{c|cc|}
\cline{2-3}
\multicolumn{1}{l|}{}                            & \multicolumn{2}{c|}{\textbf{Variability Significance {[$\sigma$]}}}             \\ \hline
\multicolumn{1}{|c|}{\textbf{Half-life {[}s{]}}} & \multicolumn{1}{c|}{\textbf{12,000 mJy}} & \textbf{6000 mJy} \\ \hline
\multicolumn{1}{|c|}{750}                        & \multicolumn{1}{c|}{2.8}                 & \textless 1        \\
\multicolumn{1}{|c|}{500}                        & \multicolumn{1}{c|}{4.2}                 & 1.7                \\
\multicolumn{1}{|c|}{250}                        & \multicolumn{1}{c|}{5.8}                 & 2.0                \\
\multicolumn{1}{|c|}{100}                        & \multicolumn{1}{c|}{10.9}                & 1.6                \\
\multicolumn{1}{|c|}{25}                         & \multicolumn{1}{c|}{5.3}                 & \textless 1        \\ \hline
\end{tabular}
\caption{Significance values for a variability detection, for simulated exponential decay stellar flares with initial amplitudes of 12,000 mJy and 6000 mJy, and half-lives of 750\,s, 500\,s, 250\,s, 100\,s, and 25\,s.}
\label{tab:sigmas}
\end{table}
  
Although the pipeline's ability to detect a varying signal is currently limited to very bright objects, due to the $\mathrm{SNR}$ threshold of the cuts algorithm (as discussed in the previous section), these results highlight the potential of the classification pipeline to detect quickly decaying flares, sub-minute flares with at least two observations, in the timestreams. This will be an essential tool in detecting fast varying (sub-minute to minute) transients.

\section{Conclusions}\label{sec:conclusions}

Our improved data cuts classification pipeline enables classifications of glitches into point sources, point sources with another coincident glitch, cosmic rays, and electronic glitches from the output of the ACT glitch finder. We have an overall accuracy of 90\%, with 94\% of PS objects being correctly classified. We are able to restore high-amplitude sources that previously would have been removed from the maps and flag the sources that would have partially passed the cuts. This enables us not only to ensure that sources are not removed from the maps but also to generate a list of times and positions of transients. Such a list can be used to issue rapid alerts and also provide informative priors for map-based searches.

The pipeline selects sources that appear multiple times in a TOD. The amplitudes of such sources can be computed across scans to generate sub-minute to minute timescale light curves. We tested the recovery of simulated light curves of the stellar flares using this pipeline. TOD-specific systematics were averaged out in order to better understand the performance we could generally expect from the amplitude computation. We found that for a bright 12,000~mJy source we can detect variability with $>4 \sigma$ for a flare with half-life $<500$\,s, and for a dimmer, 6000\,mJy source the significance was ${\sim}$1.5--2\,$\sigma$ for half-lives between 100\,s and 500\,s, with no detection of longer or shorter flares. These results are from ACT PA4 and PA5, and with more sensitive instruments, like the upcoming SO \citep{ade/etal:2019, so_collaboration:2025}, the ability to determine whether a source is varying on short time scales will improve. Finally, even when a candidate transient has low SNR, or is only seen in one or two scans, we can perform forced-aperture photometry on the source location directly in the TOD to obtain a longer light curve with a baseline flux before the transient appears.

We are adapting our classification pipeline for SO. In addition to providing data cuts for the main map-making pipeline, its transient detection capabilities are highly complementary to the map-based transient searches which are being developed. In addition to the benefits mentioned above---preventing transients from being excised from the data and providing a list of transient candidates to map-based searches---it can create high time-resolution light curves (sub-minute to minute) of transients, which cannot be readily extracted from maps. This will be helpful in issuing faster transient alerts to the broader community, since current ACT map-based transient searches require multiple TODs and the creation of depth-1 maps. Additionally, the light curves retrieved using this pipeline could allow us to rapidly classify the potential transient type (e.g. stellar flares, which are short duration, versus extragalactic explosions like GRBs, which are long duration). Furthermore, our experience thus far indicates that TOD searches may perform better on bright transients in the Galactic plane where the source density is higher, and it is more challenging for map-based searches to disentangle nearby sources from each other.

Other avenues for future work include, for instance, adding more statistics to reduce misclassified glitches, particularly for a point source with another spatially coincident glitch in the focal plane. Another interesting avenue would be to create a finer electronic glitch categorization to aid in the understanding of telescope systematics. This would especially help diagnose issues that arise when building a new telescope or implementing new hardware. This could be achieved by first implementing an unsupervised clustering algorithm to determine which categories of electronic glitches arise. We focused on using summary statistics in order to have a light-weight pipeline that can be run in real time during pre-processing. Other approaches, even though they are more memory intensive, such as image classification on the focal plane and timestreams, should also be explored. Another important area for future development is to improve the initial glitch detection algorithm so that it has a lower $\mathrm{SNR}$ cut-off, since the relatively high $\mathrm{SNR}$ threshold currently being used is a limiting factor in our performance. We are currently developing methods using machine learning to better detect glitches and sources in the data, with the goal of integrating detection into the classification pipeline. Finally, we have demonstrated our ability to recover sub-minute light curves directly from the TODs, but our simple fitting method could be further improved, particularly to obtain better error estimates for the light curves.

\section*{Acknowledgements}

Support for ACT was through the U.S.~National Science Foundation through awards AST-0408698, AST-0965625, and AST-1440226 for the ACT project, as well as awards PHY-0355328, PHY-0855887 and PHY-1214379. Funding was also provided by Princeton University, the University of Pennsylvania, and a Canada Foundation for Innovation (CFI) award to UBC. ACT operated in the Parque Astron\'omico Atacama in northern Chile under the auspices of the Agencia Nacional de Investigaci\'on y Desarrollo (ANID). The development of multichroic detectors and lenses was supported by NASA grants NNX13AE56G and NNX14AB58G. Detector research at NIST was supported by the NIST Innovations in Measurement Science program. Computing for ACT was performed using the Princeton Research Computing resources at Princeton University, the National Energy Research Scientific Computing Center (NERSC), and the Niagara supercomputer at the SciNet HPC Consortium. SciNet is funded by the CFI under the auspices of Compute Canada, the Government of Ontario, the Ontario Research Fund–Research Excellence, and the University of Toronto. We thank the Republic of Chile for hosting ACT in the northern Atacama, and the local indigenous Licanantay communities whom we follow in observing and learning from the night sky. YG acknowledges support from the University of Toronto's Eric and Wendy Schmidt AI in Science Postdoctoral Fellowship, a program of Schmidt Sciences. ADH acknowledges support from the Sutton Family Chair in Science, Christianity and Cultures, from the Faculty of Arts and Science, University of Toronto, and from the Natural Sciences and Engineering Research Council of Canada (NSERC) [RGPIN-2023-05014, DGECR-2023-00180]. CS acknowledges support from the Agencia Nacional de Investigaci\'on y Desarrollo (ANID) through Basal project FB210003. 
RH acknowledges support from the NSERC Discovery Grant Program RGPIN-2018-05750 and the Connaught Fund. The Dunlap Institute is funded through an endowment established by the David Dunlap family and the University of Toronto. The authors at the University of Toronto acknowledge that the land on which the University of Toronto is built is the traditional territory of the Haudenosaunee and, most recently, the territory of the Mississaugas of the New Credit First Nation. They are grateful to have the opportunity to work in the community on this territory. 
\software{\texttt{Astropy} \citep{astropy:2013, astropy:2018, astropy:2022}, \texttt{cutslib}, \texttt{enact}\footnote{\url{https://github.com/amaurea/enact/}}, \texttt{enlib}\footnote{\url{https://github.com/ACTCollaboration/enlib}}, \texttt{Matplotlib} \citep{Hunter:2007}, \texttt{moby2}\footnote{\url{https://github.com/ACTCollaboration/moby2/}}, \texttt{modAL} \citep{modAL2018}, \texttt{NumPy} \citep{harris2020array}, \texttt{pandas} \citep{mckinney-proc-scipy-2010, reback2020pandas}, \texttt{pickle}, \texttt{pixell}\footnote{\url{https://github.com/simonsobs/pixell}}, \texttt{scikit-learn} \citep{scikit-learn, sklearn_api}, \texttt{SciPy} \citep{2020SciPy-NMeth}, \texttt{Seaborn} \citep{Waskom2021}}



\appendix

\section{Classification Results with an Ideal Cuts Algorithm}\label{sec:ideal cuts}

\begin{figure*}
\centering
  \includegraphics[width=0.95\textwidth]{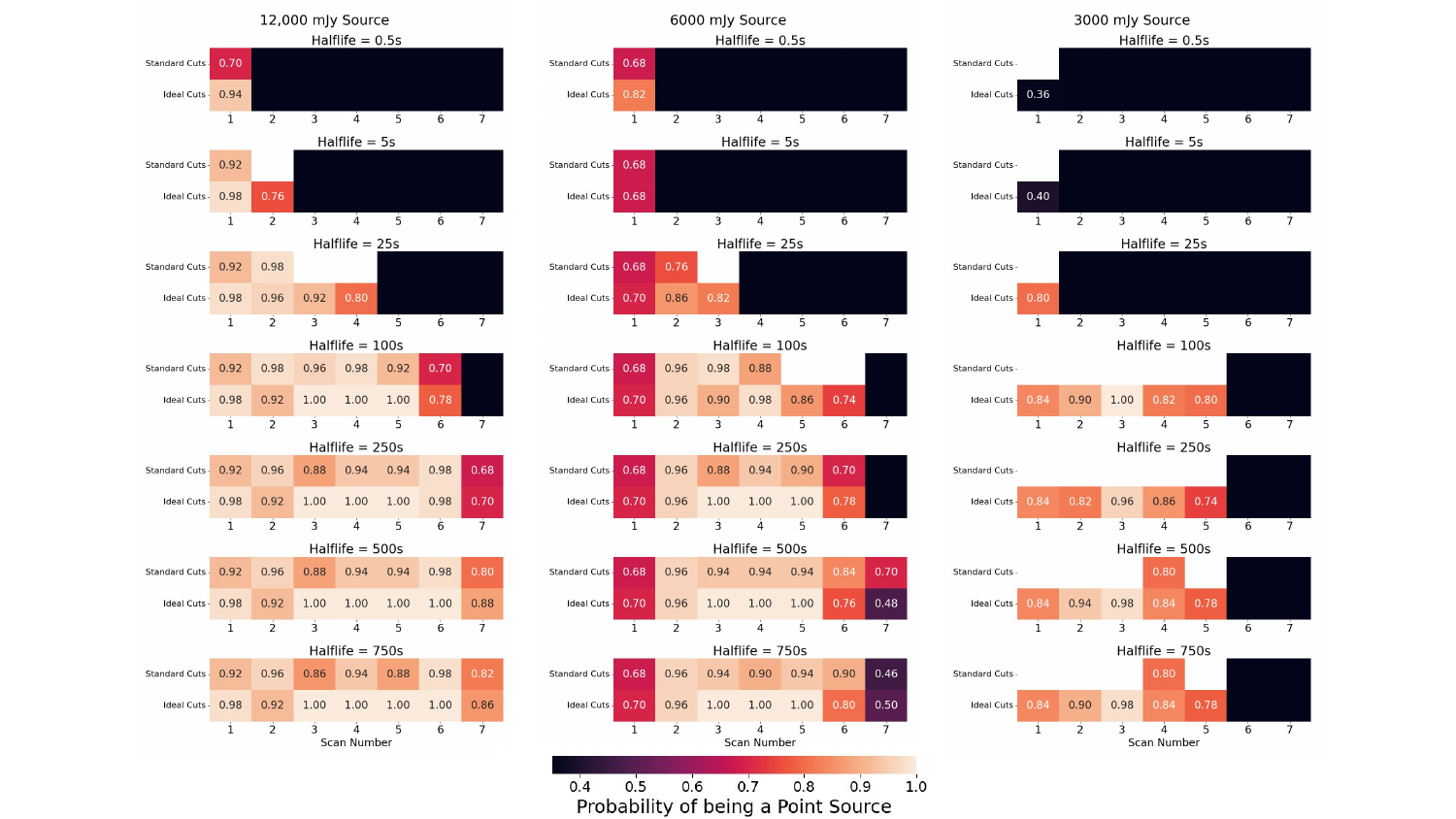}
\caption{Classification results for the simulated stellar flare with peak amplitudes of 12,000\,mJy, 6000\,mJy, 3000\,mJy, with half-lives of 0.5\,s, 5\,s, 25\,s, 100\,s, 250\,s, 500\,s, and 750\,s. For each panel, the top row is using the standard ACT cuts algorithm with our classifications and the bottom row is simulating the use of perfect cuts that flag all detectors that observe the source with an amplitude greater than 500\,mJy. Due to the telescope scanning over the same location multiple times during the TOD, we detect the source multiple times. The scan number indicates the $i^{th}$ time we detect the source. The color corresponds to the probability that the glitch is a PS according to our classification pipeline, white spaces indicate that the source is observed but that the cuts algorithm does not detect it, and black indicates that the amplitude of the source is less than 500\,mJy and is not detected. In general, we see a decrease in probability for later scans as the source's amplitude is decaying following Equation~\ref{eq:amplitude}. Note that for the 3000\,mJy injected source, fewer than four detectors were flagged for the first three scans, and thus we were not able to classify the source based on our detector conditions for glitch classification.} \label{fig: cuts and perfect cuts}
\end{figure*}

The current cuts algorithm requires an $\mathrm{SNR}$ cutoff of 10 which results in only bright sources to be flagged. In order to test if our trained random forest would be able to classify dimmer sources if they were flagged by the cuts, we utilized the stellar flare simulations to create ``ideal cuts''. Since the sources are simulated, we know the amplitudes at all times. We create modified cuts objects that flag every detector at a given time that has an amplitude greater than 500\,mJy, chosen to be just below the average noise of the detectors. We ran these modified cuts through our classification pipeline, and the results can be seen in Figure~\ref{fig: cuts and perfect cuts}. For all half-lives and amplitudes, we detected the source using these ideal cuts. This is very promising and is likely due to the importance of the focal plane statistics to our classification pipeline, as we see Figure~\ref{fig:stats importance}. In general, we see a decrease in probability for later scans as the source's amplitude is decaying following Equation~\ref{eq:amplitude}. However, this trend is not strictly followed for the first couple scans, as the ACT focal plane is a hexagon oriented such that the scan begins at one of its vertices. Thus, when the source first enters the focal plane, it only hits a few detectors. As the sky rotates through the focal plane, which is scanning at constant elevation, the source moves towards its center and hits more detectors; finally, as it exits the focal plane it again only hits a few detectors. We note that for 3000\,mJy, fewer than four detectors were flagged for the first three scans and thus were not classified, as we require $\geq$4 detectors for classification. 
 
\end{document}